\documentclass[fleqn,usenatbib,useAMS]{mnras}
\usepackage[table,xcdraw]{xcolor}
\usepackage{newtxtext}
\usepackage[utf8]{inputenc}
\usepackage[english]{babel}
\usepackage[T1]{fontenc}
\usepackage{color}
\usepackage{deluxetable}
\usepackage{amssymb,amsmath}
\usepackage{rotating}
\usepackage{graphicx}
\usepackage{multirow}
\usepackage{hyperref}
\hypersetup{colorlinks,citecolor=blue,filecolor=blue,linkcolor=blue,urlcolor=blue}
\usepackage{mathtools}
\usepackage{mathpazo}
\usepackage{graphicx}
\usepackage{lscape}
\usepackage{footnote}
\usepackage{longtable}
\usepackage{soul}
\usepackage{lastpage}
\usepackage{caption}
\usepackage{natbib}
\usepackage{lineno}
\title[Flux and color variability of Seyfert galaxies] {The Relative Contributions of Accretion Disk versus Jet to the Optical and Mid-infrared Variability of Seyfert Galaxies}

\author[Ojha Vineet, XB, LCH]{Vineet Ojha\thanks{E-mail:vineetojhabhu@gmail.com, wuxb@pku.edu.cn}$^{1}$, Xue-Bing, Wu$^{2,~1}$, Luis C. Ho$^{1,~2}$ \\
$^{1}$Kavli Institute for Astronomy and Astrophysics, Peking University, Beijing \it{100871}, China\\
$^{2}$Department of Astronomy, School of Physics, Peking University, Beijing \it{100871}, China}

\date{Accepted XXX. Received YYY; in original form ZZZ}
\pubyear{2025}
\makeatletter
\def\@journal{Preprint version 2025}
\volume{}
\pagerange{}
\pubyear{}
\def\ps@clean{
  \def\@oddfoot{\hfil}
  \def\@evenfoot{\hfil}
  \def\@oddhead{\hfil}
  \def\@evenhead{\hfil}
}
\makeatother

\begin{document}
\pagestyle{clean}
\label{firstpage}
\pagerange{\pageref{firstpage}-- \pageref{lastpage}}
\maketitle
\begin{abstract}
We performed a comprehensive analysis of flux and color variability in a redshift-matched sample of Seyfert galaxies, comprising 23 gamma-ray-detected narrow-line Seyfert 1 galaxies (gNLS1s), 190 non-gamma-ray-detected narrow-line Seyfert 1 galaxies (ngNLS1s), and 10 gamma-ray-detected broad-line Seyfert 1 galaxies (gBLS1s). Utilizing multi-band light curves from the Zwicky Transient Facility (ZTF) in g, r, and i bands, along with mid-infrared (MIR) observations in W1 and W2 bands from the Wide-Field Infrared Survey Explorer (WISE), we observed that gBLS1s exhibit more significant variability than gNLS1s, while ngNLS1s display minimal variability across both optical and MIR wavelengths. The pronounced variability in gBLS1s may be attributed to a more closely aligned jet relative to the observer’s line of sight or their comparatively lower accretion rates. In contrast, the subdued variability in ngNLS1s suggests that their flux changes are primarily driven by accretion disk instabilities. A strong correlation between optical and MIR variability amplitudes across different time scales supports the reprocessing scenario, where accretion disk emission variations are re-emitted by surrounding dust. Furthermore, our long-term color variability analysis revealed both stronger bluer-when-brighter (BWB) and redder-when-brighter (RWB) trends from the current sample, but a stronger RWB in approximately 50\%, 49\%, and 50\% of gNLS1s, ngNLS1s, and gBLS1s, respectively, in the longer side of the optical wavelength, and 55\%, 28\%, and 30\% in the MIR wavelength, strengthen the reprocessing scenario. The prevalent RWB trend observed in both optical and MIR wavelengths from the current sample on the longer time scales is likely associated with accretion disk instabilities.

\end{abstract}

\begin{keywords}
surveys -- galaxies: active -- galaxies: jets -- $\gamma$-ray-galaxies: photometry -- galaxies:
Seyfert -- Gamma-rays: galaxies
\end{keywords}

%%%%%%%%%%%%%%%%%%%%%%%%%%%%%%%%%%%%%%%%%%%%%%%%%%%%%%%%%%%%%%%%%%%%%%%%

\section{Introduction}
 \label{sec1.0}
Active galactic nuclei (AGNs) are among the most energetic extragalactic sources, with bolometric luminosities reaching up to 10$^{48}$ erg s$^{-1}$~\citep{Woo2002ApJ...579..530W}. These extraordinary objects are powered by the accretion of gas onto supermassive black holes (SMBHs) at the centers of their host galaxies, with SMBH masses ranging from 10$^{6}$ to 10$^{10}$ M$_{\odot}$. This accretion process is the primary mechanism driving their immense energy output, making AGNs the most luminous objects in the universe~\citep{Lynden-Bell1969Natur.223..690L, Rees1984ARA&A..22..471R, Koratkar1999PASP..111....1K, Bischetti2017A&A...598A.122B}. AGN models propose that, along with the central SMBH, an accretion disk and a surrounding dusty torus are present~\citep{Antonucci1993ARA&A..31..473A, Urry1995PASP..107..803U}. This structure is supported by characteristic features observed in the broadband spectral energy distribution (SED), including the big blue bump (BBB) and the mid-infrared (MIR) bump. The BBB is believed to originate from thermal emission produced by the accretion disk~\citep{Shakura1973A&A....24..337S}, while the mid-IR bump is attributed to emission from the dusty torus, as revealed by dust reverberation mapping studies~\citep{Suganuma2006ApJ...639...46S, Koshida2014ApJ...788..159K, Mandal2018MNRAS.475.5330M}.\par
Among the defining characteristics of AGNs, flux variability across the electromagnetic spectrum stands out as one of the most significant traits~\citep{Wagner1995ARA&A..33..163W, Ulrich1997ARA&A..35..445U}. Such variability, observed on timescales ranging from hours to years across wavelengths from radio to $\gamma$-ray, has been extensively utilized to investigate emitting regions, the intricate nature of the central engine, and accretion processes, phenomena~\citep[see,][]{Urry1995PASP..107..803U, Wagner1995ARA&A..33..163W, Ulrich1997ARA&A..35..445U, Czerny2008MNRAS.386.1557C, Rakshit2015MNRAS.447.2420R}. Among these, optical variability has been a main focus of study since the discovery of AGNs, as it provides crucial insights into the physics of accretion disks and jets~\citep{Wagner1995ARA&A..33..163W, Ghosh2000ApJS..127...11G, Villata2004A&A...421..103V, Gupta2008AJ....135.1384G, Rani2010MNRAS.404.1992R, Gu2011A&A...534A..59G, Ikejiri2011PASJ...63..639I, Bonning2012ApJ...756...13B, Mao2016Ap&SS.361..345M}. In addition to optical studies, the investigation of infrared (IR) variability has proven equally important. Infrared echoes, which arise in response to variations in optical continuum, provide vital information about the morphology of the dust torus and the accretion mechanisms that power the AGNs~\citep{Maune2013ApJ...762..124M}. Moreover, IR wavelengths present unique advantages over optical wavelengths, such as (1) the IR continuum being unaffected by the presence of strong emission lines in low-redshift AGNs, and (2) reduced sensitivity to dust extinction. Together, multiwavelength variability studies across the optical and IR bands provide a comprehensive understanding of the central regions and the processes shaping AGN emission. In particular, changes in the brightness of AGNs are often accompanied by spectral variations, which can be explored through color-magnitude or spectral index–magnitude correlations~\citep[e.g., see ][]{Ciprini2003A&A...400..487C}. Examining spectral variability is particularly important, as it can help to pinpoint the origin of brightness fluctuations, even when the analysis is confined to the optical and IR bands~\citep{Vagnetti2003ApJ...590..123V}.\par
Optical variability studies on large samples of luminous AGNs in optical wavelength have resulted in various interesting relationships between the amplitude and timescale of variability with physical parameters of AGN such as black hole mass, Fe~{\sc ii} strength,  accretion rate, bolometric luminosity, redshift, etc.~\citep[e.g., ][]{1996ApJ...463..466D, 2004ApJ...601..692V, Kelly2009ApJ...698..895K, MacLeod2010ApJ...721.1014M, Meusinger2011A&A...525A..37M, Rakshit2017ApJ...842...96R, Li2018ApJ...861....6L}, providing new insights into the accretion process in AGNs. However, it has been limited to a few sources only for the low luminous part of AGNs such as narrow-line Seyfert 1 galaxies (NLS1s), characterized based upon the flux ratio of forbidden line $[O~{III}]_{\lambda5007}$ with Balmer line (H$\beta$) $[O~{III}]_{\lambda5007}/H\beta$ $<$ 3, and narrow width of Balmer emission line with full width at half maximum (FWHM) $<$ 2000 km s$^{-1}$~\citep{Osterbrock1985ApJ...297..166O, Goodrich1989ApJ...342..908G}. For instance, a few studies on radio-loud and radio-quiet NLS1s have been performed on the intra-night timescale~\citep{Paliya2013MNRAS.428.2450P, Kshama2017MNRAS.466.2679K, Ojha2022JApA...43...25O, Ojha2022MNRAS.514.5607O, Ojha2024MNRAS.529L.108O} but very limited studies are available on longer time scales from days to year~\citep{Young1999MNRAS.304L..46Y, Miller2000NewAR..44..539M, Klimek2004ApJ...609...69K, Doroshenko2006ARep...50..708D, Maune2013ApJ...762..124M}. Similarly, scarce studies are available for NLS1s in the IR band~\citep{Jiang2012ApJ...759L..31J}. On the other hand, NLS1s have a higher accretion rate, strong Fe~{\sc ii} emission, stronger X-ray variability, steeper X-ray spectra, lower black hole mass, and lower optical variability compared to their broad-line counterparts, namely broad-line Seyfert 1 galaxies~\citep[BLS1s; ][]{Leighly1999ApJS..125..297L, 2001A&A...372..730V, Grupe2004ApJ...606L..41G, Zhou2006ApJS..166..128Z, Xu2012AJ....143...83X, Ojha2020ApJ...896...95O}. Despite the intrinsic differences between NLS1s and BLS1s, gamma-ray-detected Seyfert galaxies, including both NLS1s and BLS1s, are of particular interest because they exhibit optical variability on short timescales, similar to that observed in blazars~\citep{Ojha2022MNRAS.514.5607O}. This suggests a significant contribution from relativistic jets in these sources, raising the question whether the observed optical and MIR variability arises primarily from jet activity or from thermal processes in the accretion disk.\par
Although several studies have explored the variability of Seyfert galaxies, key gaps remain in our understanding of their multi-wavelength variability characteristics. For example,~\citet{Rakshit2017ApJ...842...96R} investigated long-term optical variability of the redshift-matched sample of NLS1s and BLS1s using Catalina Real-Time Transient Survey (CRTS) data, and~\citet{Rakshit2019MNRAS.483.2362R} investigated MIR flux and color variability in NLS1s using data from the Wide-Field Infrared Survey Explorer (WISE). However, a comprehensive and systematic investigation of flux and color variability from optical to MIR wavelengths, specifically comparing gamma-ray-detected NLS1s (gNLS1s), non-gamma-ray NLS1s (ngNLS1s), and gamma-ray-detected BLS1s (gBLS1s) is still lacking. Such a study is crucial for disentangling the relative contributions of jet and disk emissions in these systems, thereby advancing our understanding of the physical mechanisms driving variability across the electromagnetic spectrum.
Given the availability of new catalog of NLS1s~\citep{Paliya2024MNRAS.527.7055P}, which is a factor of two times larger than the previous catalog given by~\citet{Rakshit2017ApJS..229...39R} and high cadence ($\sim$ 1 day) photometric data, available in Zwicky Transient Facility~\citep[hereafter,  ZTF;][]{Bellm2019PASP..131a8002B} for $\emph {g}$, $\emph {r}$, and $\emph {i}$  bands along-with MIR data from WISE~\citep{Wright2010AJ....140.1868W}, strongly motivate us to do a systematic study of flux and color variability for a sample of gNLS1, ngNLS1s, and gBLS1s across optical to MIR wavelengths. Therefore, for the optical study of gNLS1s, ngNLS1s, and gBLS1s, we used the ZTF~\citep[see, ][]{Bellm2019PASP..131a8002B, Masci2019PASP..131a8003M}, which is an optical time-domain survey utilizing the 48-inch Schmidt Telescope at Palomar Observatory. It is equipped with a 6k$\times$6k mosaic CCD  camera that provides a field of view (FoV) of 47 deg$^{2}$ with a readout time of 8 sec~\citep{Graham2019PASP..131g8001G} and scans the northern sky in $\emph {g}$, $\emph {r}$, and $\emph {i}$  bands with an average cadence up to 1 day.\par
Additionally, for the MIR study of the current samples, we utilized data from WISE, which surveys the entire sky in four MIR bands $\emph {W1}$, $\emph {W2}$, $\emph {W3}$, and $\emph {W4}$ centered at 3.4 $\mu$m, 4.6 $\mu$m, 12 $\mu$m, and 22 $\mu$m, respectively, with angular resolutions of 6.1, 6.4, 6.5, and 12 arcseconds, respectively. WISE has a FoV of 47 arcmin$^{2}$ and completes a full-sky scan approximately every six months. During each visibility window, it continuously observes sources for at least one day, occasionally longer, with a cadence of 90 minutes. Observations were conducted during two primary phases: (i) the WISE Full Cryogenic and Near-Earth Object WISE (NEOWISE) Post-Cryo Missions (2010–2011), which covered three visibility windows in all four bands, with photometric data archived in the “AllWISE Multi-Epoch Photometry (MEP) Database”; and (ii) the NEOWISE Reactivation Mission~\citep[NEOWISE-R; ][]{Mainzer2014ApJ...792...30M}, spanning 2014–2024, which focused on the $\emph {W1}$ and $\emph {W2}$  bands across eight visibility windows, with data stored in the “NEOWISE-R single exposure (L1b) source table.\par
The format of this paper is as follows. In Sect.~\ref{section_2.0}, we describe our sample selection and data compilation. Sect.~\ref{sec_3.0} provides details of our methodologies used for the analysis of the data. The results of the current work, followed by discussion, are given in Sect.~\ref{sec_4.0}. The conclusions of the current work are described in Sect.~\ref{sect_5.0}.
 
\section{Sample selection}
\label{section_2.0}
\begin{figure}
    \centering
    \includegraphics[width=1.0\linewidth,height=1.0\linewidth]{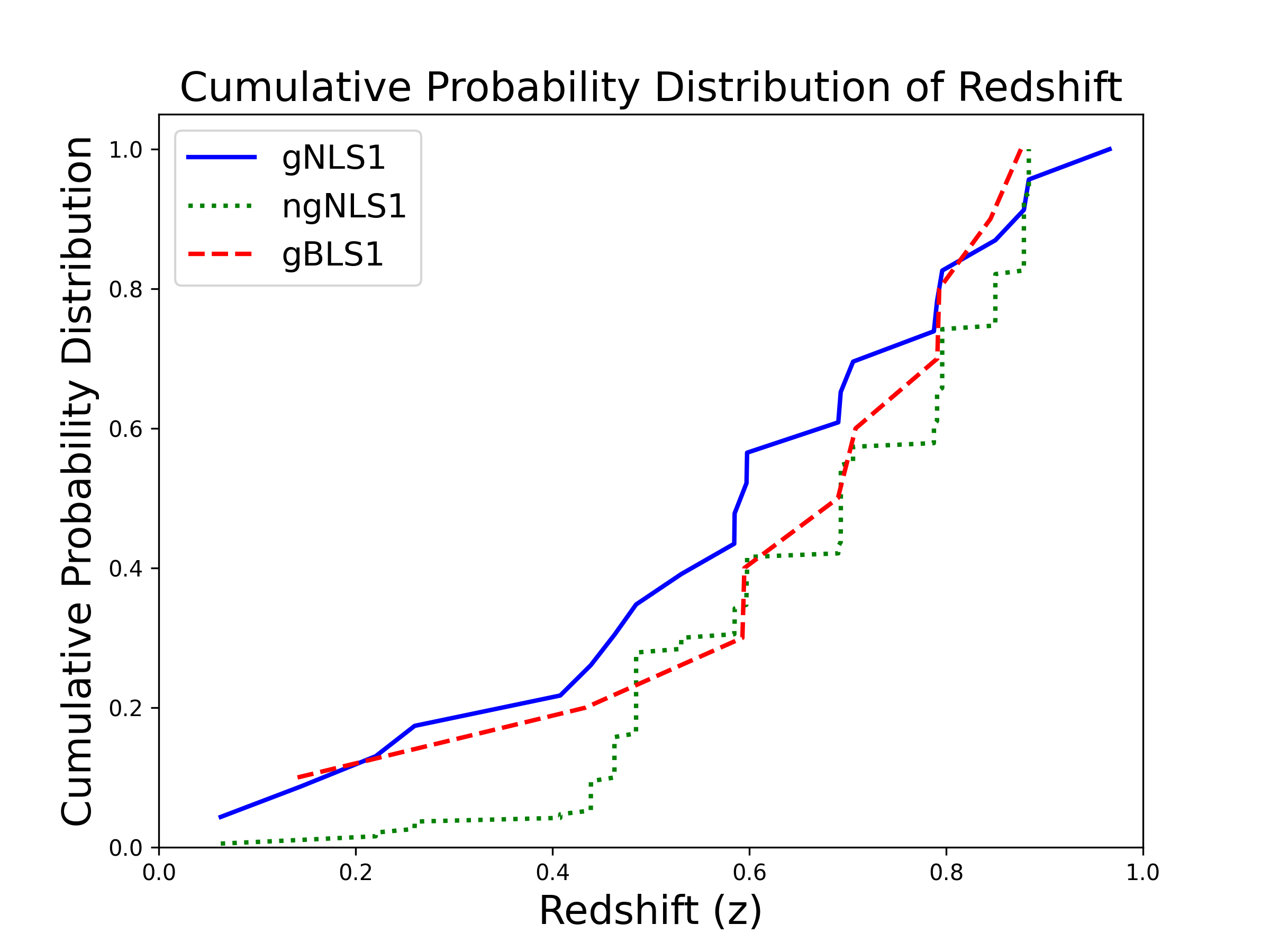}
    \caption{Cumulative probability distribution of rest-frame redshifts for our samples of 23 gNLS1s (solid blue), 190 ngNLS1s (dotted green), and 10 gBLS1s (dashed red).}
    \label{fig: histo_redshift}
\end{figure}

For this study, we used the most recent catalog of NLS1 galaxies, comprising 22,656 sources~\citep{Paliya2024MNRAS.527.7055P}. To construct a sample of ngNLS1s, we first compile all known gNLS1s identified by the {\it Fermi} Large Area Telescope (LAT) from the literature, including the 21 sources listed in Table 1 of~\citet{Paliya2024MNRAS.527.7055P}. This compilation yielded a total of 25 gNLS1s. Applying a redshift cut of $z \leq 1$ to ensure adequate signal-to-noise ratios (S/N) for the variability analysis using data from ZTF and WISE reduced the gNLS1 sample to 23 sources. In parallel, a control sample of 45 gBLS1s was selected from~\citet{Paliya2024MNRAS.527.7055P}, also subject to the same redshift constraint of $z \leq 1$ to maintain consistency in data quality across samples. Thus, finally, we identified 22,635 ngNLS1s, 23 gNLS1s, and 45 gBLS1s. Next, we prepared redshift-matched samples to enable a fair comparison. The resulting matched samples consist of 23 gNLS1s, 190 ngNLS1s, and 10 gBLS1s whose cumulative probability distribution for rest-frame redshifts is presented in Fig.~\ref{fig: histo_redshift}. Finally, a Kolmogorov–Smirnov (K-S) test performed on the redshift-matched samples yielded a null hypothesis probability ($P_{\text{null}}$) of approximately 21\% for gNLS1s versus ngNLS1s and 28\% for gNLS1s versus gBLS1s.\par
Out of different subsamples of Seyfert galaxies used in the current work, gNLS1s and gBLS1s can exhibit hour-like time-scale variability, which can create an obstacle in studying the true spectral behavior of these sources in the absence of simultaneous multi-band data, both in optical and MIR wavebands. Therefore, for the color variability study, we selected sources that have quasi-simultaneous observations within 30 minutes in different bands with at least five data points. To make it consistent, we have applied the same criterion of quasi-simultaneous observations to the sample of ngNLS1s. This has resulted in a sample of 17 gNLS1s, 128 ngNLS1s, and 10 gBLS1s for the $\emph {g-r}$ color study and 10 gNLS1s, 74 ngNLS1s, and 6 gBLS1s for the $\emph {r-i}$ color study in the optical wavelength. Similarly, samples of 22 gNLS1s, 176 ngNLS1s, and 10 gBLS1s were obtained for the $\emph {W1-W2}$ color study at the mid-infrared wavelength.

\subsection{Optical photometric data from ZTF}
For the redshift-matched samples of 23 gNLS1s, 190 ngNLS1s, and 10 gBLS1s, we retrieved ZTF photometric data for the $\emph {g}$, $\emph {r}$, and $\emph {i}$  band in the 22$^{nd}$ ZTF public data release~\citep[see][]{Masci2019PASP..131a8003M} within a search radius of one arcsec to the target position from NASA/IPAC IR Science Archive (IRSA\footnote{https://irsa.ipac.caltech.edu/Missions/ztf.html}) using the application program interface (API). To eliminate any suspicious measurements that could mimic spurious variability, such as those arising from the independent calibration of ZTF light curves across different fields and CCD quadrants for the same filter~\citep[e.g., see][]{2021AJ....161..267V}, we select only the light curve for a given filter corresponding to the observation ID with the maximum number of data points. Furthermore, to ensure the quality of the photometric data, we selected data points with catflags = 0, as recommended in the ZTF Science Data System Explanatory Supplement\footnote{https://irsa.ipac.caltech.edu/data/ZTF/docs/ztf explanatory
supplement.pdf} and the public data release notes\footnote{https://irsa.ipac.caltech.edu/data/ZTF/docs/releases/dr20/ztf release notes dr20.pdf}, removed outliers by applying a 3$\sigma$ clipping on the entire data set and also excluded poor quality data points with uncertainties greater than 10\% in magnitude. The precautions and filtering above enable us to determine small-magnitude variations in the generated $\emph {g}$, $\emph {r}$, and $\emph {i}$  band light curves. This has resulted in samples of 20 gNLS1s, 186 ngNLS1s, and 10 gBLS1s for the ZTF $\emph {r}$ band. The total number of sources with available data in the ZTF $\emph {g}$, $\emph {r}$, and $\emph {i}$  bands for different classes of Seyfert galaxies is tabulated in Table~\ref{ZTF_variability_table}. The obtained lightcurves of Seyfert galaxies have magnitude data points at a cadence of as high as $\sim$ 1 day for sources in their field with a maximum time baseline of $\sim$ 2000 days. Representative light curves in the $\emph {r}$ band for one of the gNLS1-J032441.28$+$341045.1 ($\emph{left}$), ngNLS1-J221541.64$+$301435.7 ($\emph{middle}$) and gBLS1-
J095649.88$+$251516.1 ($\emph{right}$) from the current sample are shown in Fig.~\ref{fig: OP_flux_variability}.

\subsection{Mid-infrared data from WISE}
For the redshift-matched samples of 23 gNLS1s, 190 ngNLS1s, and 10 gBLS1s, we retrieved MIR data from the MEP database and the NEOWISE-R single exposure (L1b) source database from the NASA/IPAC IR Science Archive\footnote{https://irsa.ipac.caltech.edu/Missions/wise.html}, giving us data from 2010 to 2024 with a gap of approximately 3 years between 2011 and 2014 due to the depletion of the solid hydrogen cryogen used to cool the instruments $\emph {W3}$ and $\emph {W4}$. Next, we limited our MIR study on the current sample to the $\emph {W1}$ and $\emph {W2}$  bands only due to the availability of the data for the $\emph {W3}$ and $\emph {W4}$ bands between 2010 and 2011, with many photometric points marked as ‘null’. Furthermore, both databases were screened to exclude bad photometric measurements, adopting the criteria (i) the reduced value of $\chi^{2}$ for the single exposure profile fit must be below 5 for both the $\emph {W1}$ and $\emph {W2}$  bands (that is, w1rchi2 $<$ 5 and w2rchi2 $<$ 5). (ii) The number of points spread function (PSF) components used in the profile fit for a source should be fewer than 3 (nb $<$ 3). (iii) Single-exposure images must have the highest quality (qi$\_$fact = 1), be free from known artifacts (cc$\_$flags = ‘0000’), and not involve active d-eblending (na = 0). (iv) The number of photometric data points available in the $\emph {W1}$ and $\emph {W2}$  bands to be $\geq$ 5. (v) Also removed outliers by applying a 3$\sigma$ clipping on the whole data set.\par
With the above constraints, we arrived at final samples of 22 gNLS1s, 178 ngNLS1s, and 10 gBLS1s for their study of MIR flux variability. The total number of sources with available data in the WISE $\emph {W1}$ and $\emph {W2}$  bands for different classes of Seyfert galaxies is tabulated in Table~\ref{ZTF_variability_table}. Furthermore, upon visually inspecting the generated light curves for the $\emph {W1}$ and $\emph {W2}$  bands, we observed that many light curves included photometric points separated by approximately 11 seconds in consecutive WISE epochs. Given that the WISE orbital period is roughly 1.5 hours, we averaged such photometric points with their nearest neighbors within this time frame to reconstruct the light curves for the entire sample. Representative light curves for one of the gNLS1-163915.80$+$412833.7 (\emph{left}), ngNLS1-J145758.08$+$454119.2 (\emph{middle}), and gBLS1-J103632.97$+$220312.2 (\emph{right}) from the current sample for the WISE $\emph {W1}$ band ($\emph {top panel}$) and $\emph {W2}$ band (\emph {bottom panel}) are shown in Fig.~\ref{fig: IR_flux_variability}.

\section{Methodology for analysis}
\label{sec_3.0}
Before proceeding with the analysis further, we first corrected the observed data of each source for the galactic interstellar reddening and absorption foreground, using the extinction correction (A$_{V}$) values provided by NASA’s NED, following the method of~\citep{Schlafly2011ApJ...737..103S}. The extinction-corrected $\emph {r}$-band light curves for a gNLS1, ngNLS1, and gBLS1 from the current sample are shown in Fig.~\ref{fig: OP_flux_variability}.

\subsection{Optical flux variability}
To confirm the presence/absence of variability in the final samples of gNLS1s, ngNLS1s, and gBLS1s (see Table~\ref{ZTF_variability_table}) using the ZTF $\emph {g}$, $\emph {r}$, and $\emph {i}$  band light curves, we have used the standard $\emph F^{\eta}$-test~\citep[see,][]{Goyal2012A&A...544A..37G}. A detailed explanation for this test is demonstrated in our previous papers~\citep[][and references therein]{Ojha2022MNRAS.514.5607O, Ojha2024MNRAS.529L.108O}.\par

In summary, the following~\citet{Goyal2012A&A...544A..37G}, $F^{\eta}$-test can be written as

\begin{equation}
\begin{aligned}
 \label{eq.fetest}
F_{Sy}^{\eta} = \frac{\sigma^{2}_{(Sy)}} { \eta^2 \langle \sigma_{Sy}^2 \rangle} 
\end{aligned}
\end{equation}

where $\sigma^{2}_{(Sy)}$ is the variance and $\langle \sigma_{Sy}^2 \rangle=\sum_ {i=1}^{N}\sigma^2_{i,~err}(Sy)/N$ is the mean square (formal) rms errors of the i$^{th}$ data points in the light curves of the target Seyfert galaxy, and $N$ is the number of observations. A necessary requirement of this test is to use the correct rms errors on the photometric data points. However, in determining the instrumental magnitudes of the target in the ZTF data reduction pipeline, the Image Reduction and Analysis Facility (IRAF\footnote{http://iraf.noao.edu/}) software package and the Dominion Astrophysical Observatory Photometry (DAOPHOT II\footnote{http://www.astro.wisc.edu/sirtf/daophot2.pdf}) algorithm were used, which typically underestimate magnitude errors by a factor ranging from 1.3 to 1.75, as noted by previous studies~\citep{Sagar2004MNRAS.348..176S, Bachev2005MNRAS.358..774B}. To correct for this underestimation, a value derived from 262 intra-night AGN monitoring sessions by~\citet{Goyal2013JApA...34..273G} using a ground-based telescope of a meter class has been adopted. This correction is crucial for accurate photometric analysis, particularly for the rms errors of the photometric data points. In their study,~\citet{Goyal2013JApA...34..273G} estimated the best fit value of $\eta$ to be 1.54$\pm$0.05, which accounts for the systematic underestimate of the magnitude errors by the data reduction software and ensures the reliability of the variability analysis. Therefore, in the current study, $\eta = 1.54$ is considered due to the high precision of this estimate.
To ensure a robust assessment of variability in the optical light curves of Seyfert galaxies, we employed the \emph{F}-test, which compares the observed variance in the light curve to the expected variance due to measurement uncertainties, scaled by a factor $\eta$ in this work to use the rms errors of the photometric data points accurately. We adopted a stringent significance level of $\alpha = 0.01$, corresponding to a confidence level of 99\%, to minimize the probability of a Type I error (i.e., false identification of variability when none exists). In statistical terms, setting $\alpha = 0.01$ means that there is only a 1\% chance of incorrectly rejecting the null hypothesis of no variability, thus ensuring that any detected variability is supported with high confidence. If a galaxy light curve is classified as variable at this threshold, we are 99\% confident that the observed variability is intrinsic and not due to random fluctuations in the data points. The test statistic denoted as $F_{Sy}^{\eta}$ and defined in Eq.~\ref{eq.fetest}, was calculated for each light curve and compared with its critical F-value, $F_{cr}$, obtained from the \emph{F} distribution for the appropriate degrees of freedom at the chosen significance level. The critical value $F_{cr}$ depends on the degrees of freedom (here equal to the number of data points in the observed light curve) in the numerator and denominator. A galaxy is deemed variable (V) if its computed value of the \emph{F}-test satisfies the condition $F_{Sy}^{\eta} \geq F_{cr}$, thus rejecting the null hypothesis and confirming the presence of statistically significant variability at the confidence level of 99\%.\par
Furthermore, to estimate the level of brightness variations in the above samples, we have estimated (i) the peak-to-peak amplitude of flux variability ($\psi_{\mathrm{pp}}$) following the definition given by~\citet{Heidt1996A&A...305...42H} and (ii) the fractional variability amplitude (F$_{\mathrm{var}}$) introduced by~\citet{Vaughan2003MNRAS.345.1271V} for all data sets.\par

Following the definition given by~\citet{Heidt1996A&A...305...42H}, the $\psi_{\mathrm{pp}}$ can be written as 

\begin{equation} 
\label{eq.Amp_pp}
\hspace{1.5cm} \psi_{\mathrm{pp}}= \sqrt{({Sy_{Amax}}-{Sy_{Amin}})^2-2\sigma^2}
\end{equation} 

where $Sy_{Amax}$ and $Sy_{Amin}$ are the maximum and minimum amplitude values in the light curve and $\sigma^2 = \eta^2\langle\sigma^2_{Sy}\rangle$ is the mean square (formal) rms errors of individual data points scaled with the scaling factor $\eta = 1.54$, adopted from~\citet{Goyal2013JApA...34..273G}. Here, the precession of the light curve $\sqrt{ \langle \sigma_{i,~err}^2 \rangle}$ is considered for the error associated with $\psi_{\mathrm{pp}}$.\par
Furthermore, to estimate the normalized intrinsic source variance free from the contribution expected from measurement errors, we follow Equation~\ref{eq.Fvar} introduced by~\citet{Vaughan2003MNRAS.345.1271V}

\begin{equation}
\label{eq.Fvar}
\hspace{2.5cm} F_{\mathrm{var}} = \sqrt{\frac{Sy^2 - \overline{\sigma_{\mathrm{syerr}}^2}}{\overline{\mu}^2}}.
\end{equation}

where, $\overline{\sigma_{\mathrm{syerr}}^2}$ is the mean square error and is given as follows:
\[
\overline{\sigma_{\mathrm{syerr}}^2} = \frac{1}{N} \sum_{i=1}^{N} \sigma_{\mathrm{syerr},i}^2, ~~and~~
Sy^2 = \frac{1}{N-1} \sum_{i=1}^{N} (\mu_i - \overline{\mu})^2
\]
is the sample variance for the flux measurements $N$, $\mu_i$. The associated error with $F_{\mathrm{var}}$ can be computed following the given equation.

\begin{equation}
\label{eq.err_Fvar}
\sigma(F_{\mathrm{var}}) = \sqrt{\left( \sqrt{\frac{1}{2N}} \frac{\overline{\sigma_{\mathrm{syerr}}^2}}{\overline{\mu}^2 F_{\mathrm{var}}} \right)^2 + \left( \sqrt{\frac{\overline{\sigma_{\mathrm{syerr}}^2}}{N}} \frac{1}{\overline{\mu}} \right)^2}.
\end{equation}

For each gNLS1, ngNLS1, and gBLS1, we have tabulated the duration of the light curve, the number of data points in the light curve, the variability statistics, based on the test $F^{\eta}$, $\psi_{\mathrm{pp}}$ and $F_{\mathrm{var}}$ in Table~\ref{tab:OP_IR_variability}.

\subsubsection{Duty cycle of variability}
AGNs do not necessarily exhibit flux variations during each epoch of their monitoring. Hence, it is more suitable to evaluate the duty cycle (DC) as the ratio of the time intervals when variability is observed to the total observing time rather than simply considering the fraction of variable objects. Following the definition proposed by~\citet{Romero1999A&AS..135..477R} and further utilized by~\citet{Stalin2004JApA...25....1S} for intra-night variability, we computed the DC for the current sample of Seyfert galaxies using the following expression

\begin{equation} 
\hspace{2.5cm} DC = 100\frac{\sum_{m=1}^n S_m(1/\Delta T_m)}{\sum_{m=1}^n (1/\Delta T_m)} 
\hspace{0.1cm}{\rm per~cent} 
\label{eq.DC} 
\end{equation} 

here, $\Delta T_m = \Delta T_{m,~observed}(1+$z$)^{-1}$ represents the observed duration of the $m^{th}$ monitoring session after applying the redshift correction to the source. For the $m^{th}$ session, $S_m$ is assigned a value of 1 in Eq.~\ref{eq.DC} only when variability is observed, otherwise $S_m$ is set to 0.\par
In Table~\ref{ZTF_variability_table}, we have tabulated the computed DC along with the mean and median values of $\psi_{\mathrm{pp}}$ and $F_{\mathrm{var}}$ for the current sample of Seyfert galaxies across the $\emph {g}$, $\emph {r}$ and $\emph {i}$  bands. It is important to note that when calculating the mean amplitude ($\overline{\psi_{\mathrm{pp}}}$) and the median amplitude (${\psi_{\mathrm{pp}}}$) for a sample, only light curves classified as variables were considered.   

\subsection{Mid-infrared flux variability}
 To quantitatively study the MIR variability of the current sample, we calculated the intrinsic amplitude of the variability ($V_{m}$) commonly used by several authors to study the MIR variability~\citep[e.g., see][]{Rakshit2019MNRAS.483.2362R, Anjum2020MNRAS.494..764A, Wang2020RAA....20...21W}. The $V_{m}$ is the variance of the observed magnitudes after correcting for associated measurement errors, which can be calculated following~\citet{Ai2010ApJ...716L..31A} and adopting the formalism used in~\citet{Sesar2007AJ....134.2236S}.
 
\begin{equation}
\Sigma = \sqrt{\frac{1}{n-1} \sum_{k=1}^{N}(msy_k - \langle msy \rangle)^2},
\end{equation}
where $msy_k$ is the magnitude observed at the $k^{th}$ point and $\langle msy \rangle$ is the weighted average. The $V_m$ can be expressed as:
\[
V_m = 
\begin{cases} 
\sqrt{\Sigma^2 - p_{err}^2}~~~~, & \text{if } \Sigma > p_{err} \\
~~~~~~~~0~~~~~~~~~~~~~~, & \text{otherwise}
\end{cases}
\]
where $p_{err}$ is calculated including the systematic error ($p_{s}$) with the individual error ($p_{k}$) as follows:

\begin{equation}
p_{err}^2 = \frac{1}{N} \sum_{k=1}^{N} p_k^2 + p_s^2.
\end{equation}

The uncertainty associated with $V_m$ can be computed following the given equation

\begin{equation}
\sigma_{V_m} = \frac{1}{\sqrt{\Sigma^2 - p_{\text{err}}^2}} \sqrt{ \Sigma^2 \, \sigma_{\Sigma}^2 + p_{\text{err}}^2 \, \sigma_{p_{\text{err}}}^2 }
\end{equation}

considering that the data points are independent and normally distributed, the uncertainty in $\Sigma$ can be approximated as:
\[\sigma_{\Sigma} \approx \frac{\Sigma}{\sqrt{2(n - 1)}} \]

here, systematic errors of 0.024 mags and 0.028 mags are used for the $\emph {WISE}$ $\emph {W1}$ and $\emph {W2}$  bands, respectively, as reported in~\citet{Jarrett2011ApJ...735..112J}. Since the current sample has a redshift range between $0.01 < z < 1.00$, therefore, spurious variability may occur if we do not correct $V_{m}$ for the redshift of the source, thus we calculated the rest frame $V_{mz}$ by multiplying $V_{m}$ by $\sqrt{(1+z)}$. Here, we have considered a Seyfert galaxy to be variable if its computed value of $V_{m}$ is found to be $\geq$ 0.1. The $V_{mz}$ for the individual source and its variability status are tabulated in Table~\ref{tab:OP_IR_variability} for the $\emph {W1}$ and $\emph {W2}$ bands and calculated values of $V_{mz}$, along with their associated uncertainties, for the samples of gNLS1s, ngNLS1s, and gBLS1s in the $\emph {W1}$ and $\emph {W2}$ bands are presented in Table~\ref{ZTF_variability_table}.

\subsection{Correlation between optical and mid-infrared variability amplitudes}
\label{correlation_var_amp}
\begin{figure*}
    \begin{minipage}[]{1.0\textwidth}
   \includegraphics[width=0.33\textwidth,height=0.25\textheight,angle=00]{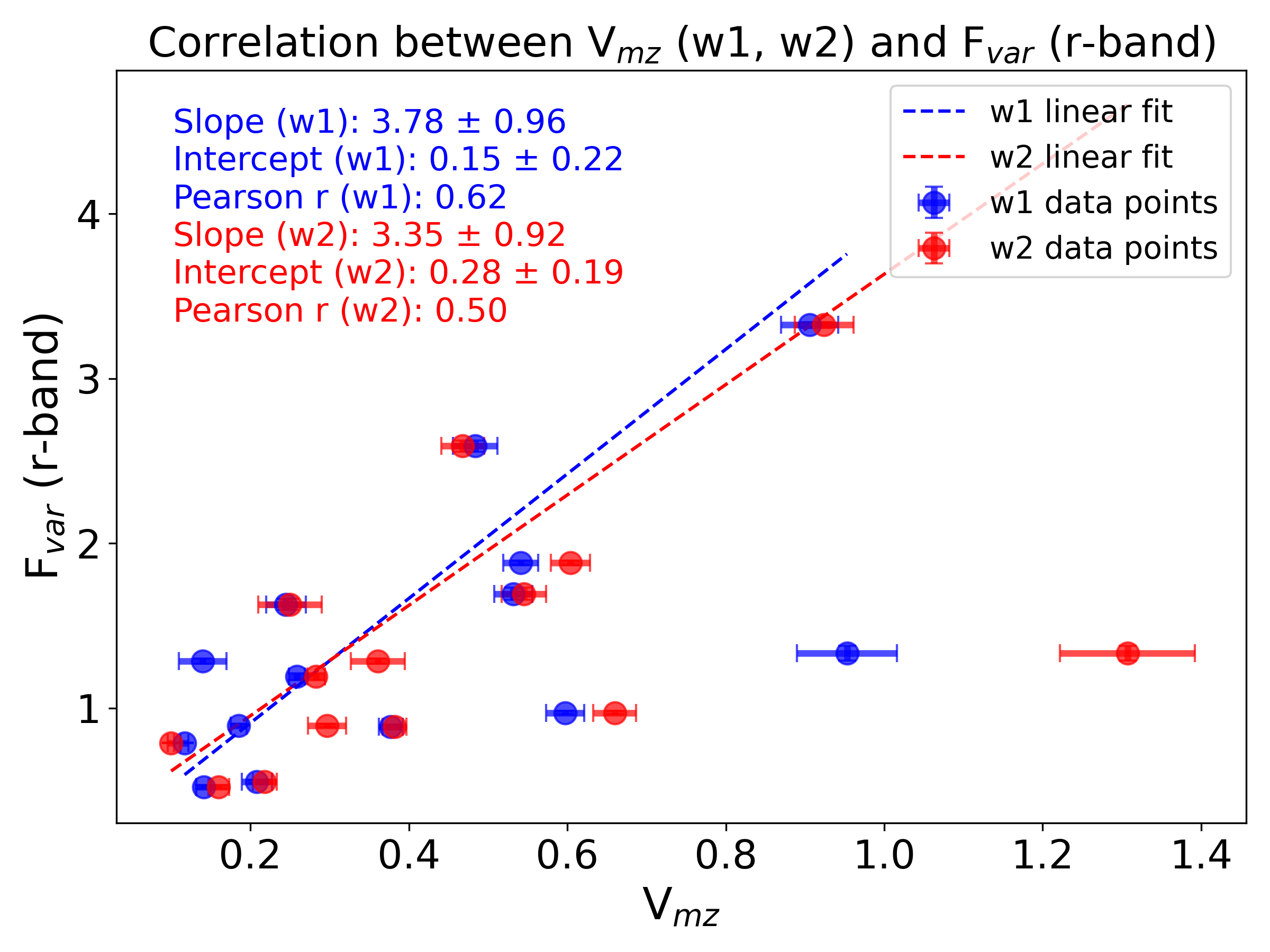} %left   
   \includegraphics[width=0.33\textwidth,height=0.25\textheight,angle=00]{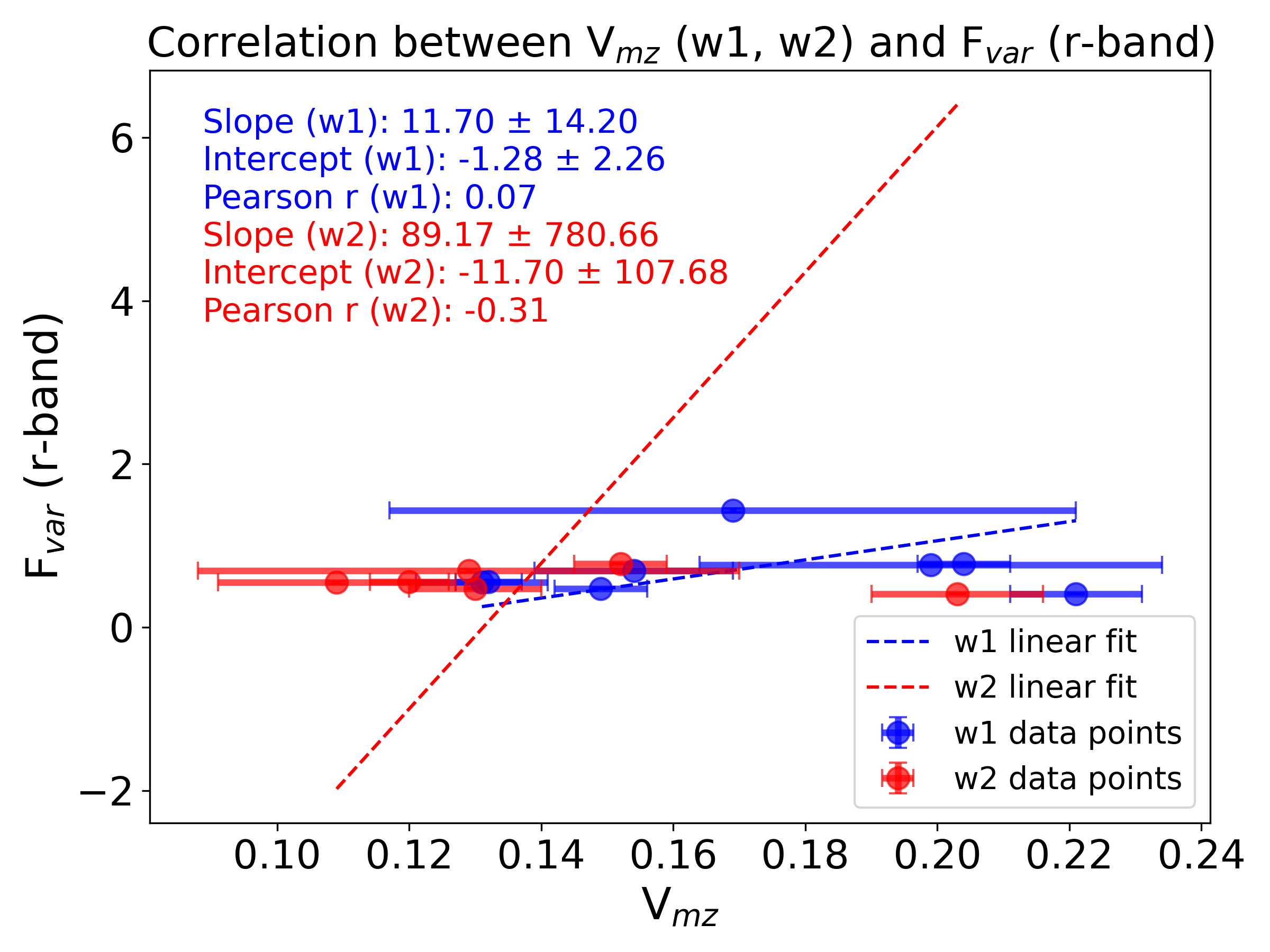} %middle
   \includegraphics[width=0.33\textwidth,height=0.25\textheight,angle=00]{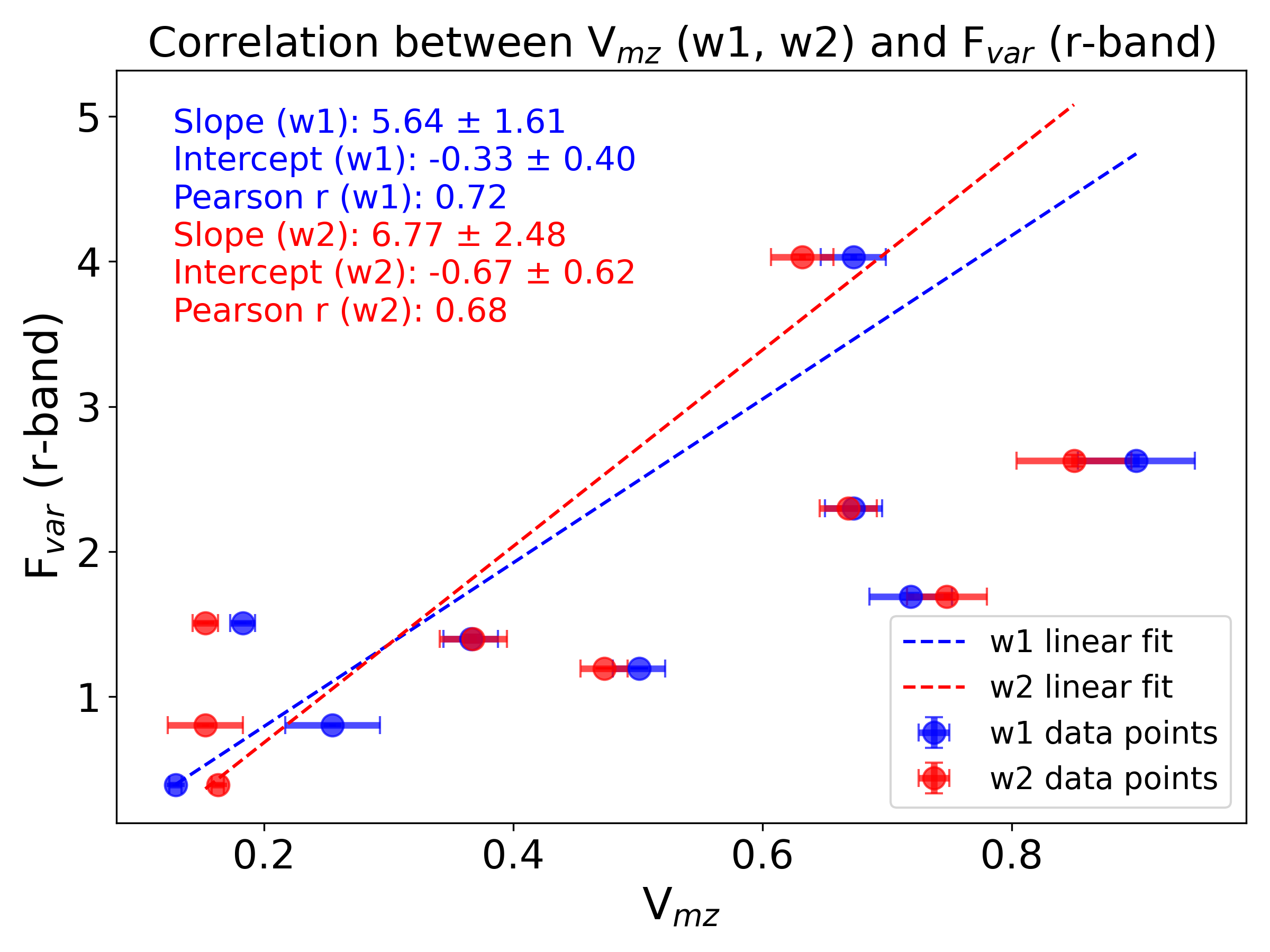} \\ %right
  \includegraphics[width=0.33\textwidth,height=0.25\textheight,angle=00]{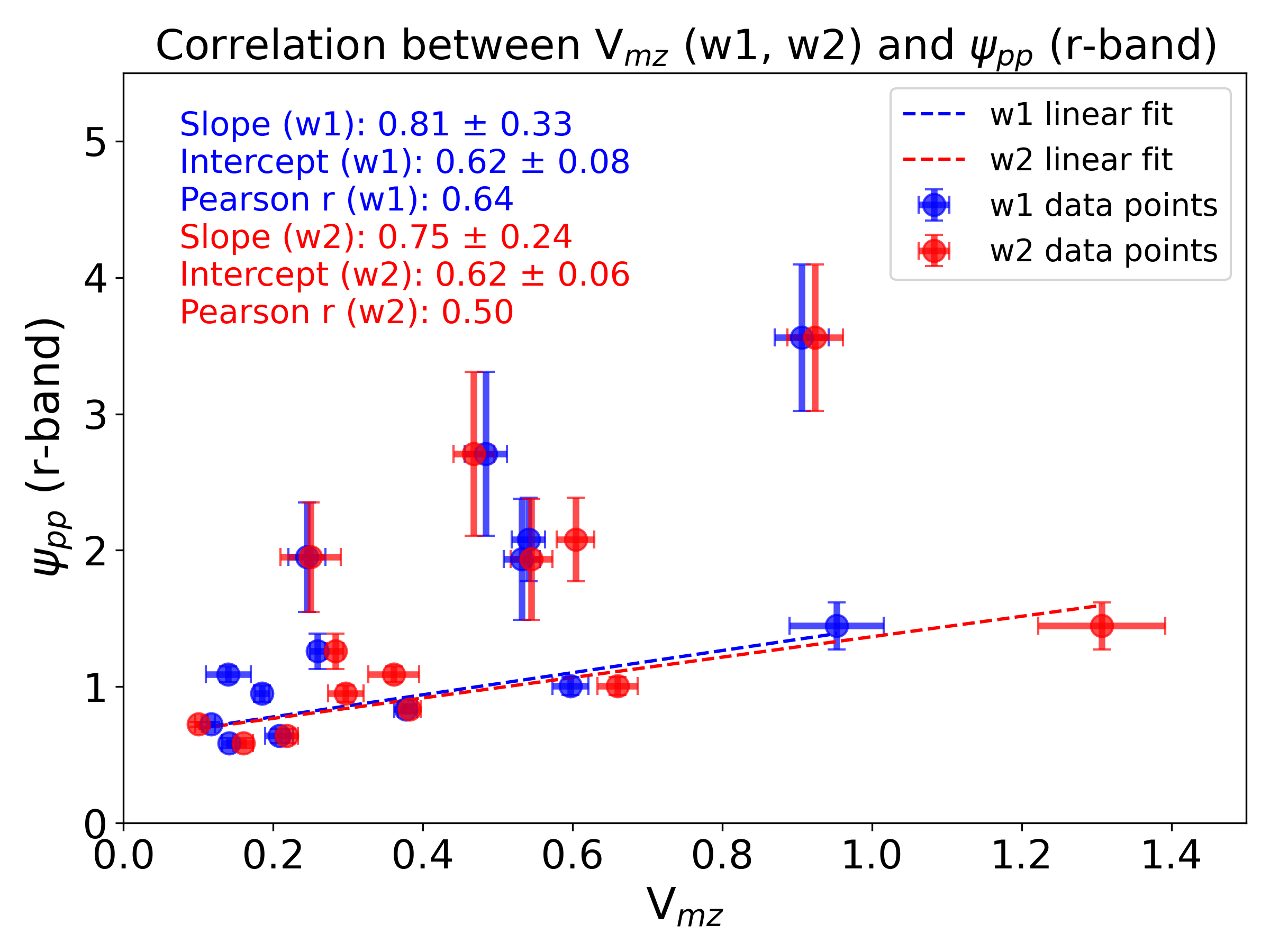} %left
\includegraphics[width=0.33\textwidth,height=0.25\textheight,angle=00]{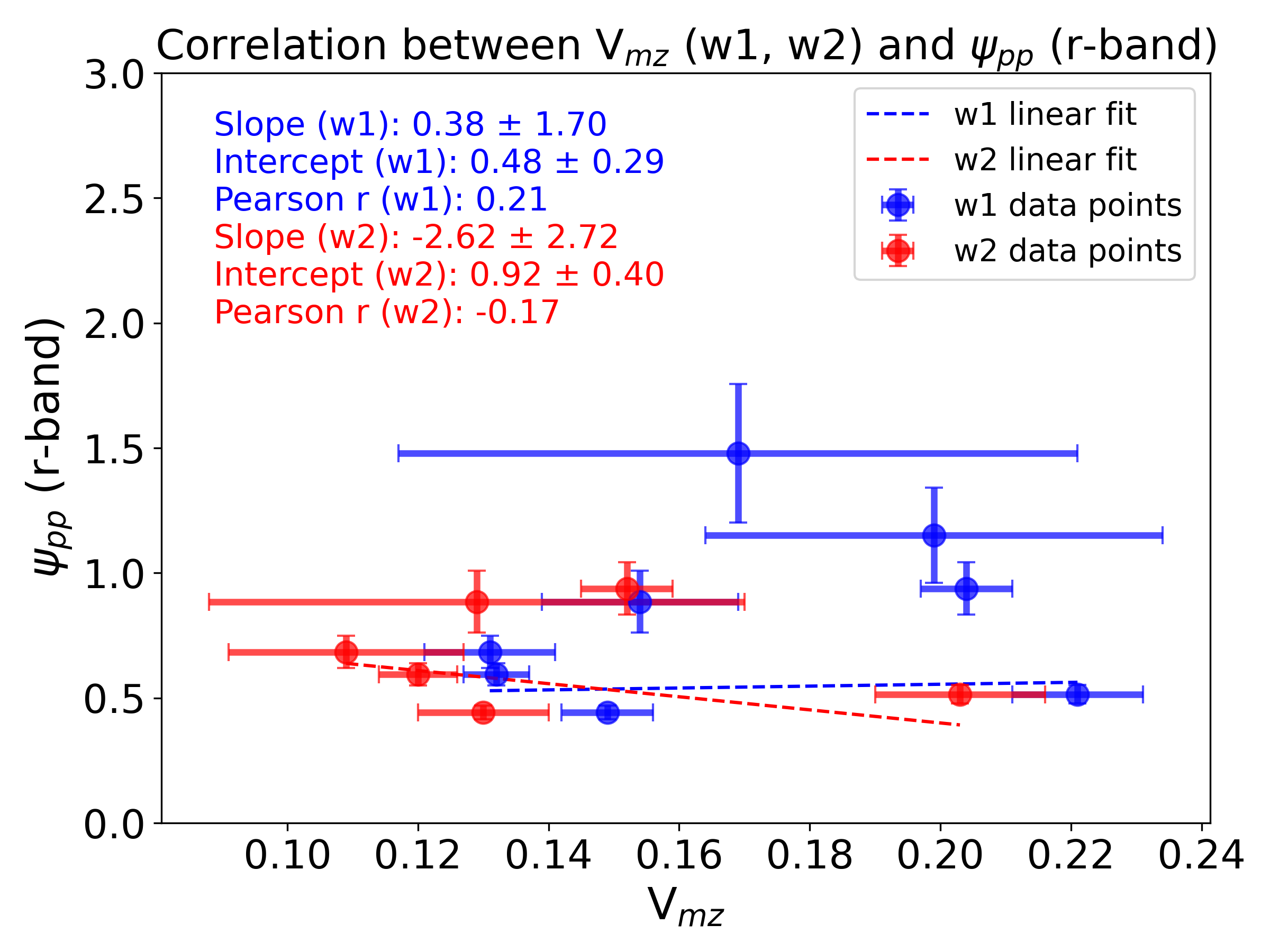} %middle
\includegraphics[width=0.33\textwidth,height=0.25\textheight,angle=00]{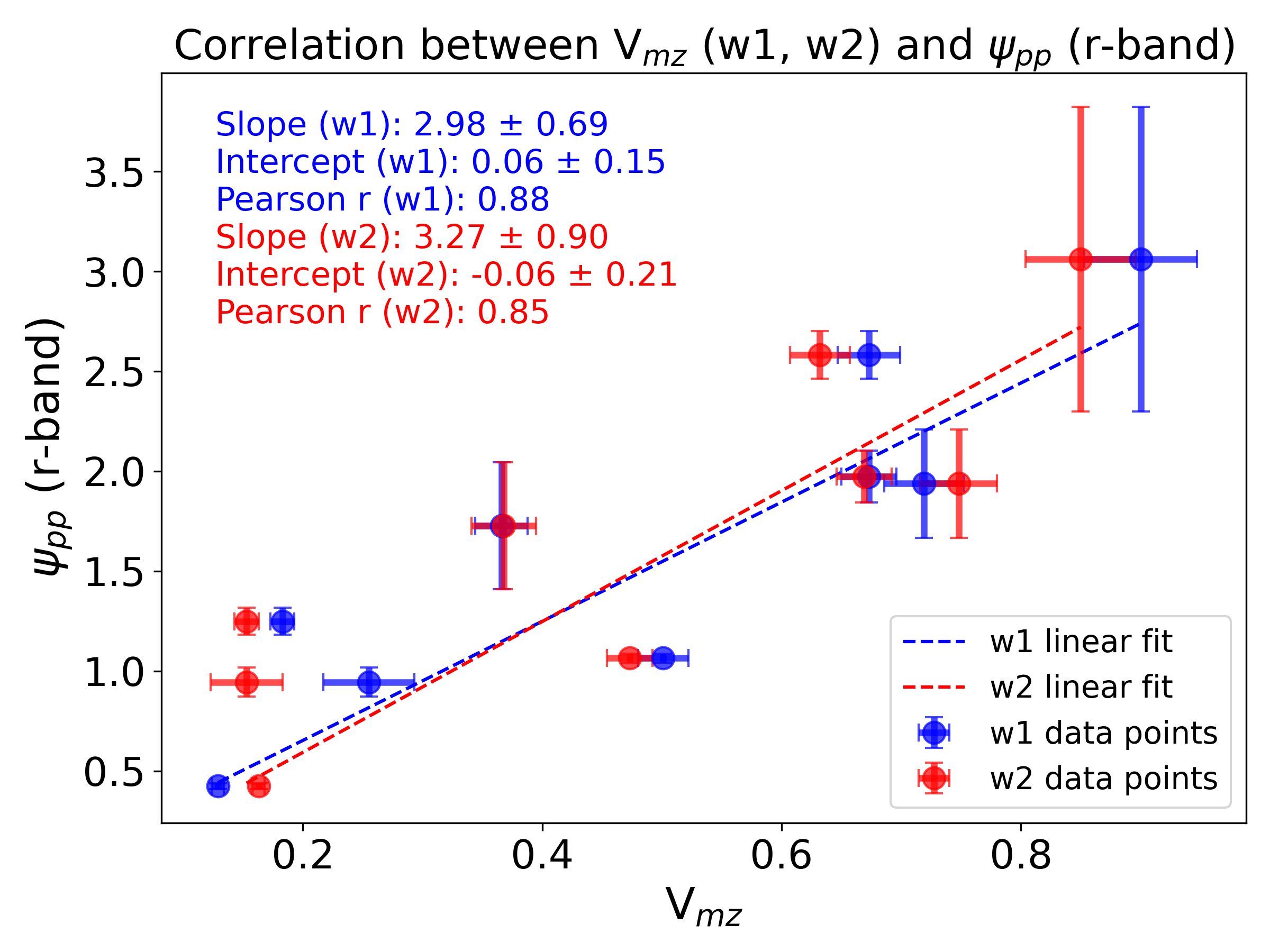} %right
    \end{minipage}
    \caption{The figure presents the correlated variability amplitude of the $\emph {r}$-band  optical light curve with the mid-infrared $\emph {W1}$ and $\emph {W2}$  band light curves for the samples of gNLS1s (\emph{left}), ngNLS1s (\emph{middle}), and gBLS1s (\emph{right}). The upper panels illustrate the correlation between the redshift-corrected variability amplitude ($V_{mz}$) and the fractional variability ($F_{\mathrm{var}}$), while the lower panels depict the correlation between $V_{mz}$ and the peak-to-peak variability amplitude ($\psi_{\mathrm{pp}}$). Note that only optically variable sources having V$_{mz} \geq$ 0.1 in the mid-infrared wavelength are considered. Blue and red dashed lines, respectively represent the best linear fit for $\emph {W1}$ and $\emph {W2}$ bands, obtained using orthogonal distance regression, accounting for uncertainties on both axes measurements. The best-fit linear regression parameters, including the slope, intercept, and Pearson correlation coefficient (Pearson r), are displayed in the upper-left corner of each panel for $\emph {W1}$ and $\emph {W2}$ bands.}
    \label{fig: correlated_optical_and_infrared_var}
\end{figure*}

It is expected that IR emission is reprocessed from optical/UV emission~\citep{Lu2016MNRAS.458..575L}, which means that variability characteristics in the optical bands will be imprinted in the IR wavelength. To investigate such a scenario in the current sample, we selected Seyfert galaxies that exhibit optical variability and have a redshift-corrected variability amplitude $V_{mz} \geq 0.1$ in the MIR. For the r-band light curve, the criteria are satisfied by 14 gNLS1s, 8 ngNLS1s, and 9 gBLS1s in both the $\emph {W1}$ and $\emph {W2}$ bands. However, for the $\emph {W2}$ band, only 6 ngNLS1s meet the criteria, while 8 ngNLS1s meet the criteria for the $\emph {W1}$ band. Using this sample, we constructed correlation plots between $V_{mz}$ and both $F_{\mathrm{var}}$ and $\psi_{\mathrm{pp}}$. A representative plot showing the correlated variability amplitudes of the $\emph {r}$ band optical light curve with the MIR $\emph {W1}$ and $\emph {W2}$ bands for gNLS1s (left), ngNLS1s (middle), and gBLS1s (right) is presented in Fig.~\ref{fig: correlated_optical_and_infrared_var}. To examine the relationship between MIR and optical variability amplitudes, we employed orthogonal distance regression (ODR) fitting on the data points in the variability amplitude correlation plot, accounting for uncertainties on both axes. Furthermore, we calculated the Pearson rank correlation coefficient ($\rho_{r}$) to quantify the correlation between the variability amplitudes for the gNLS1s, ngNLS1s, and gBLS1s samples at both the optical and MIR wavelengths. In this plot, the upper panels illustrate the correlation between $V_{mz}$ and $F_{\mathrm{var}}$, while the lower panels show the correlation between $V_{mz}$ and $\psi_{\mathrm{pp}}$. For the $\emph {r}$ band light curve, the results of the correlation analysis, such as slope, intercept, Pearson r for the $\emph {W1}$ and $\emph {W2}$ bands, are displayed in the upper left corner of each panel in Fig.~\ref{fig: correlated_optical_and_infrared_var}.

\begin{figure*}
    \begin{minipage}[]{1.0\textwidth}
    \includegraphics[width=0.34\textwidth,height=0.25\textheight,angle=00]{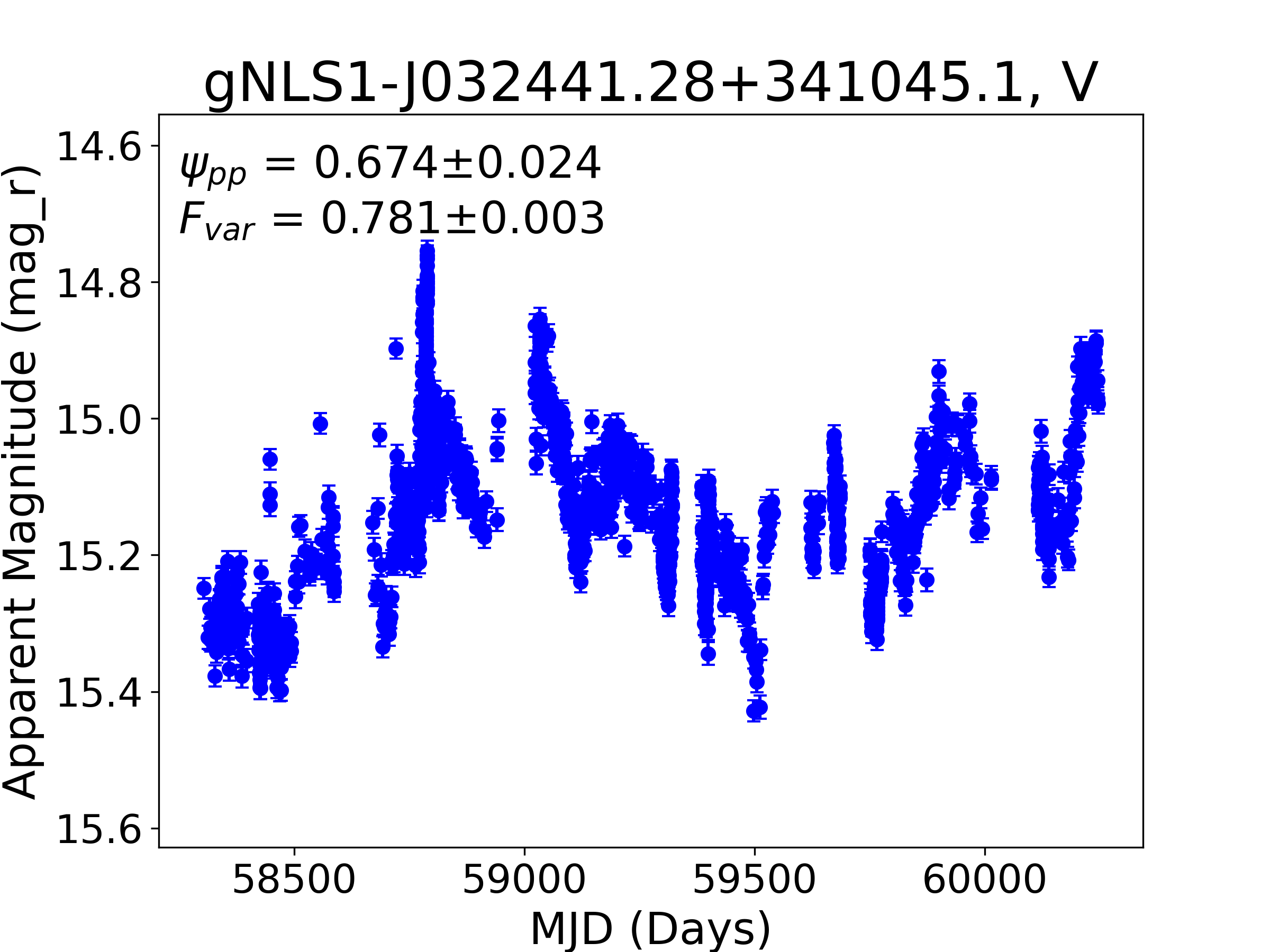}  %left
   \includegraphics[width=0.34\textwidth,height=0.25\textheight,angle=00]{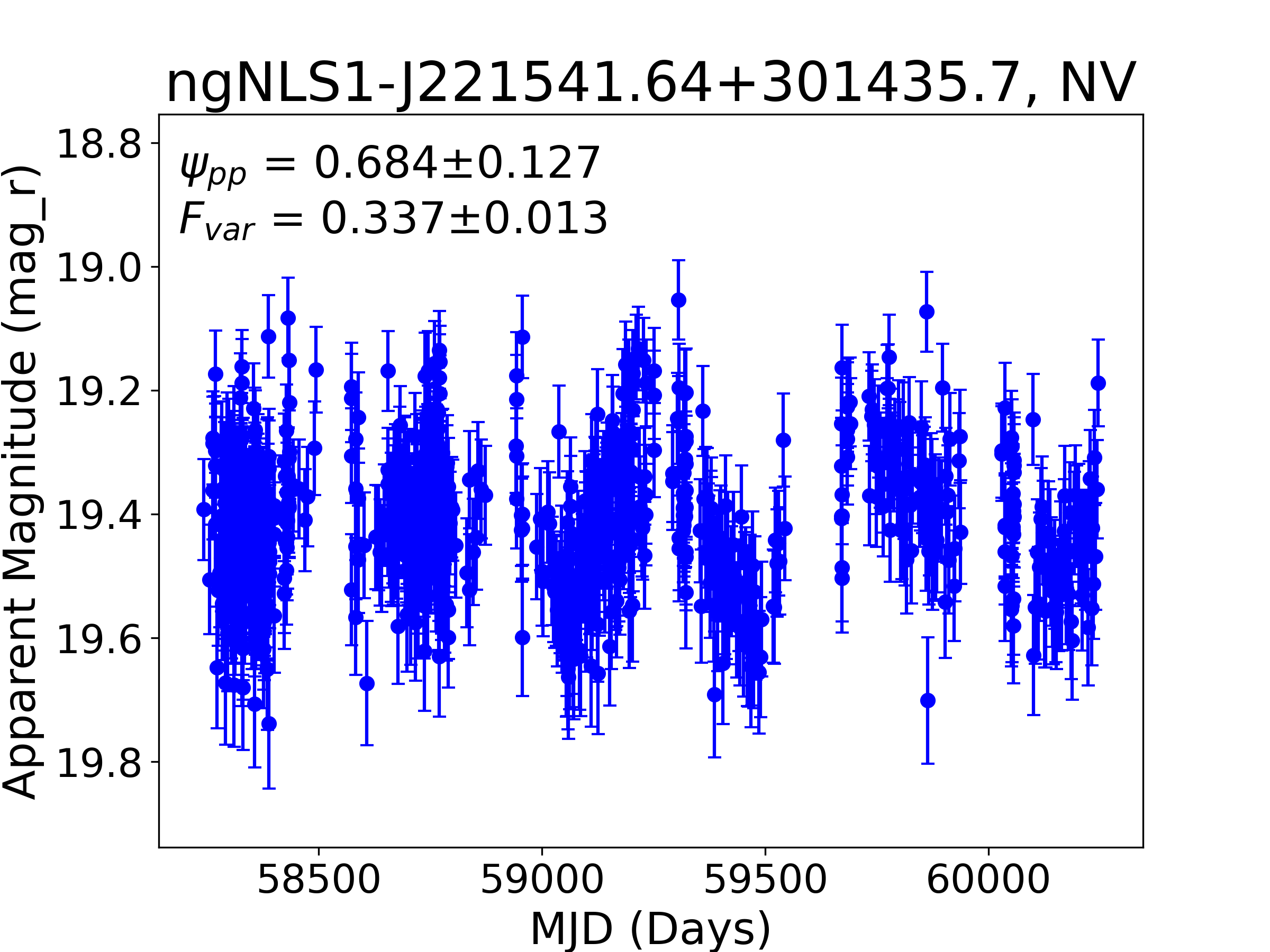} %middle   
   \includegraphics[width=0.34\textwidth,height=0.25\textheight,angle=00]{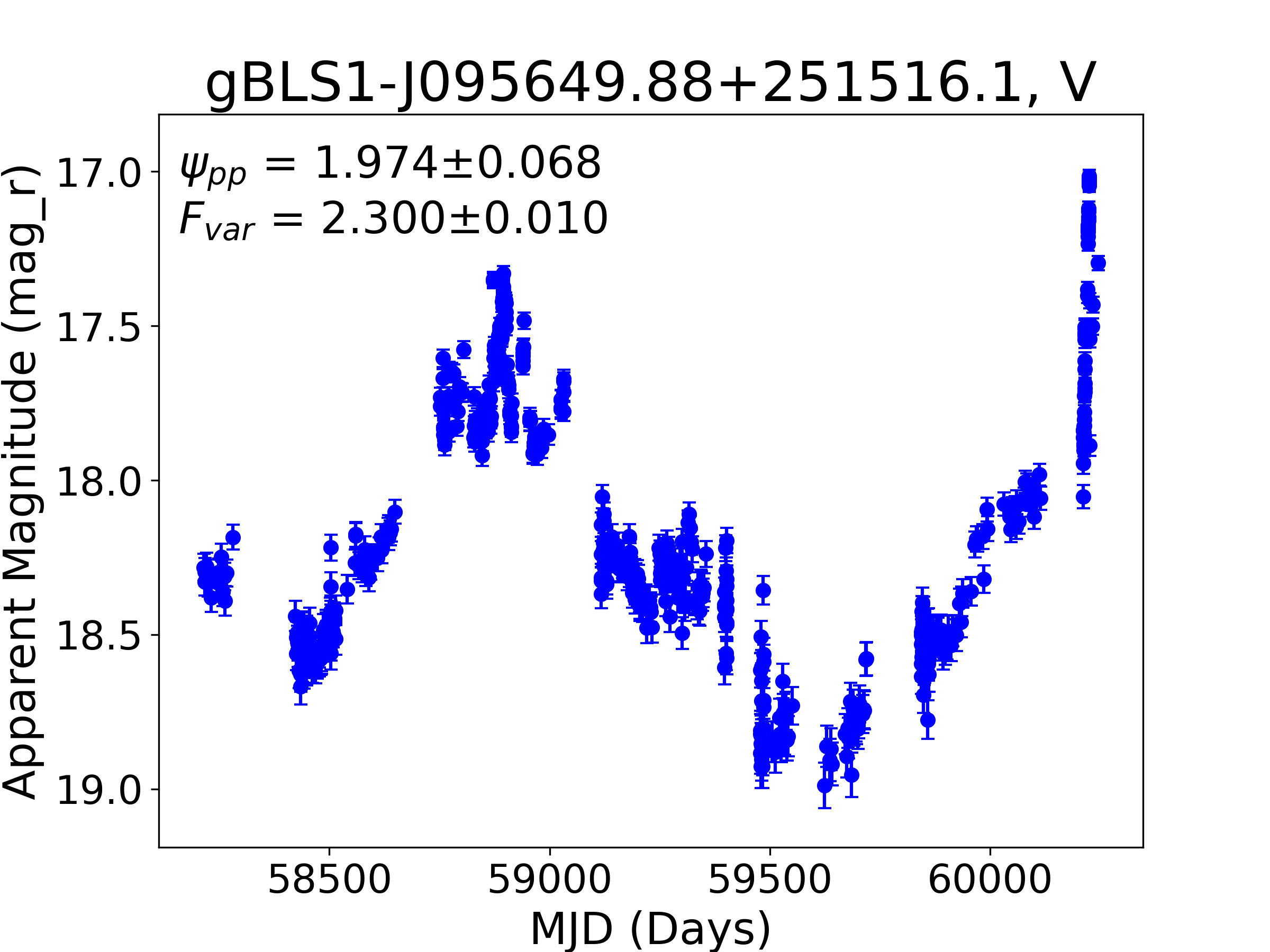}  %right
    \end{minipage}
    \caption{The $\emph {r}$-band  ZTF light curves for one of the gNLS1-J032441.28$+$341045.1 (\emph{left}), ngNLS1-J221541.64$+$301435.7 (\emph{middle}), and gBLS1-J095649.88$+$251516.1 (\emph{right}) from the current sample, showing variability (V), non-variability (NV), and variability (V) on year-like timescale. The estimated variability parameters $\psi_{\mathrm{pp}}$ and $F_\mathrm{var}$ are displayed in the upper-left corner of each panel.}
    \label{fig: OP_flux_variability}
\end{figure*}

\begin{figure*}
    \begin{minipage}[]{1.0\textwidth}
    \includegraphics[width=0.34\textwidth,height=0.25\textheight,angle=00]{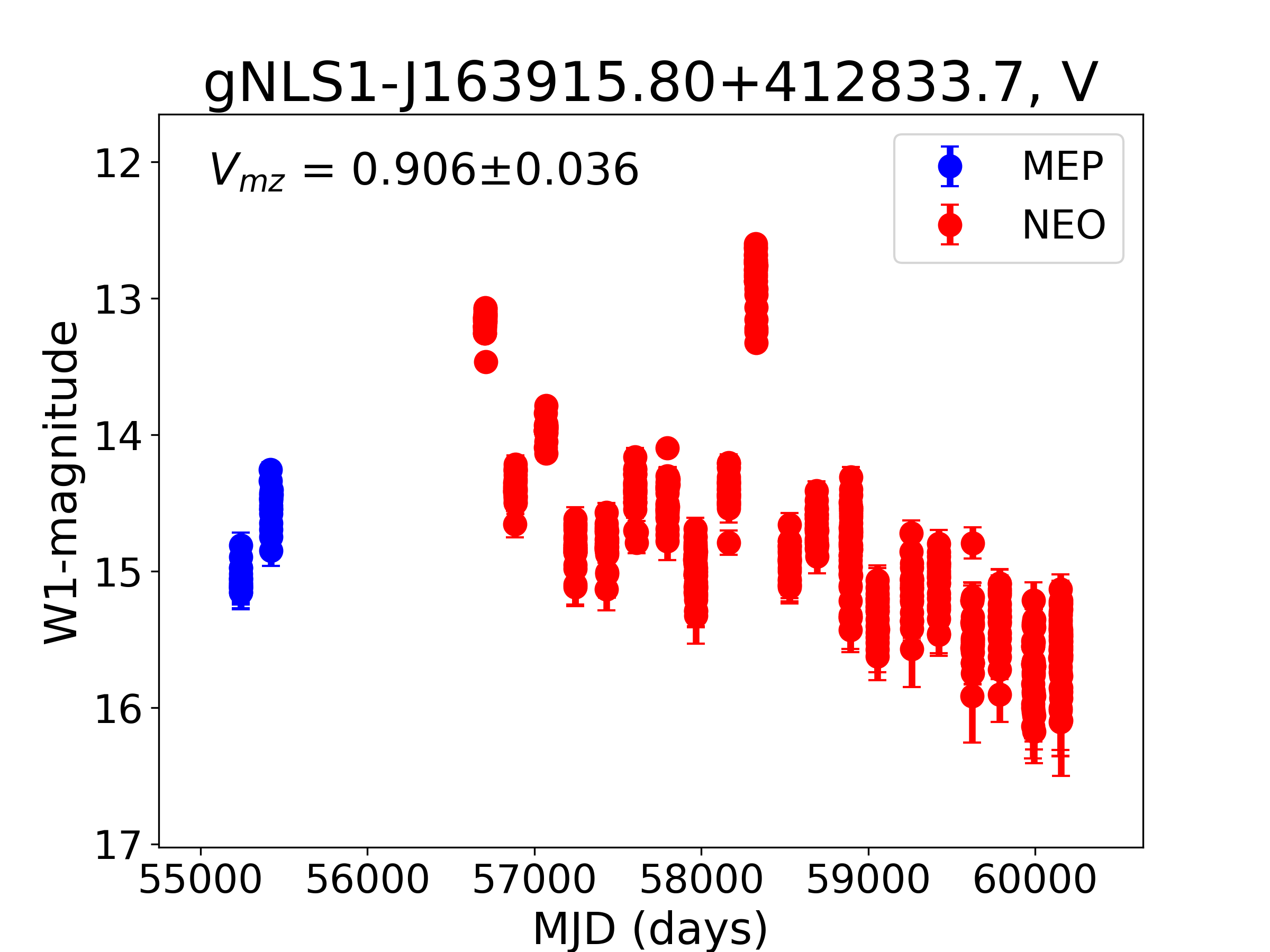}  %left
   \includegraphics[width=0.34\textwidth,height=0.25\textheight,angle=00]{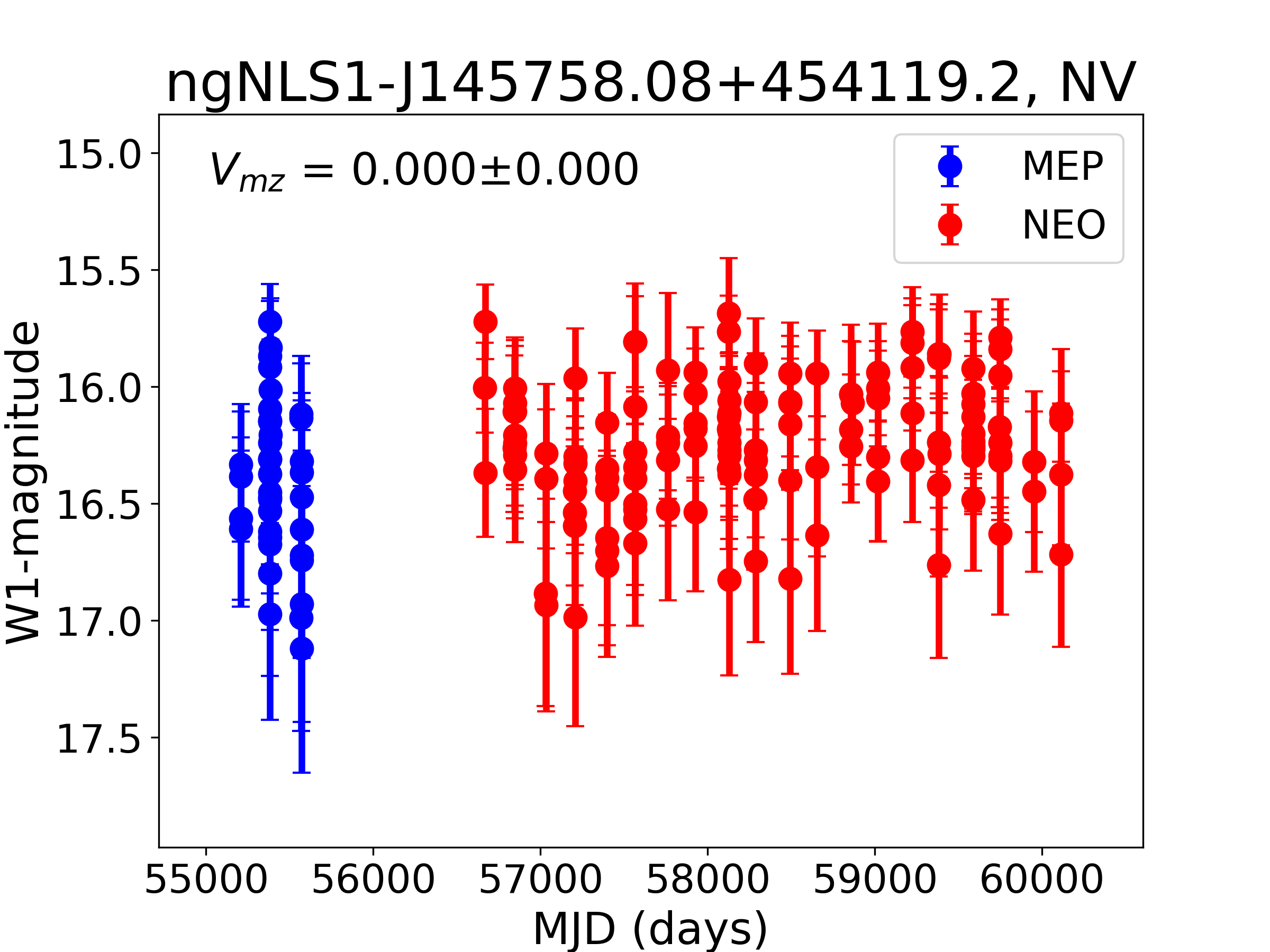} %middle   
   \includegraphics[width=0.34\textwidth,height=0.25\textheight,angle=00]{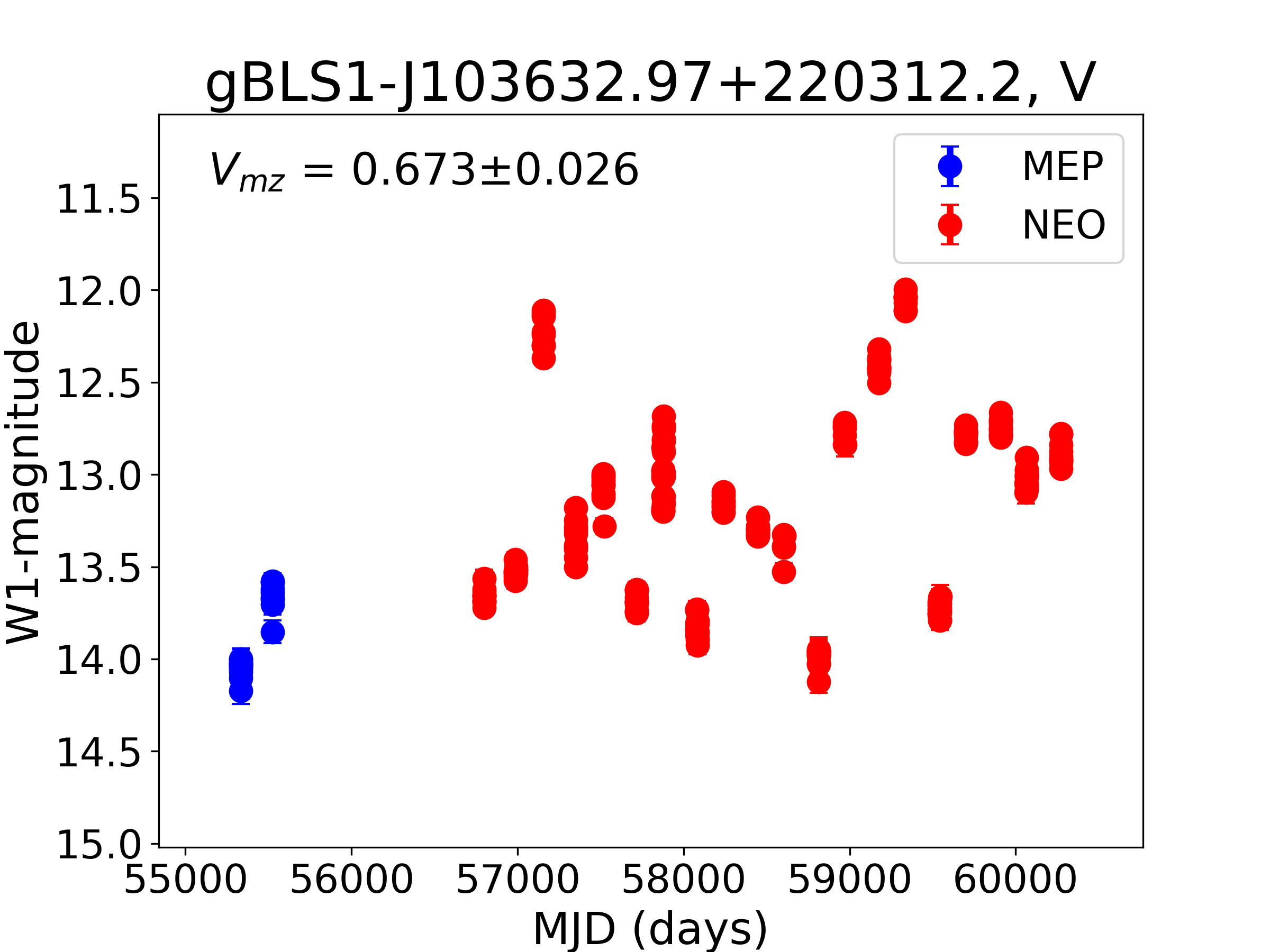}  %right
   \includegraphics[width=0.34\textwidth,height=0.25\textheight,angle=00]{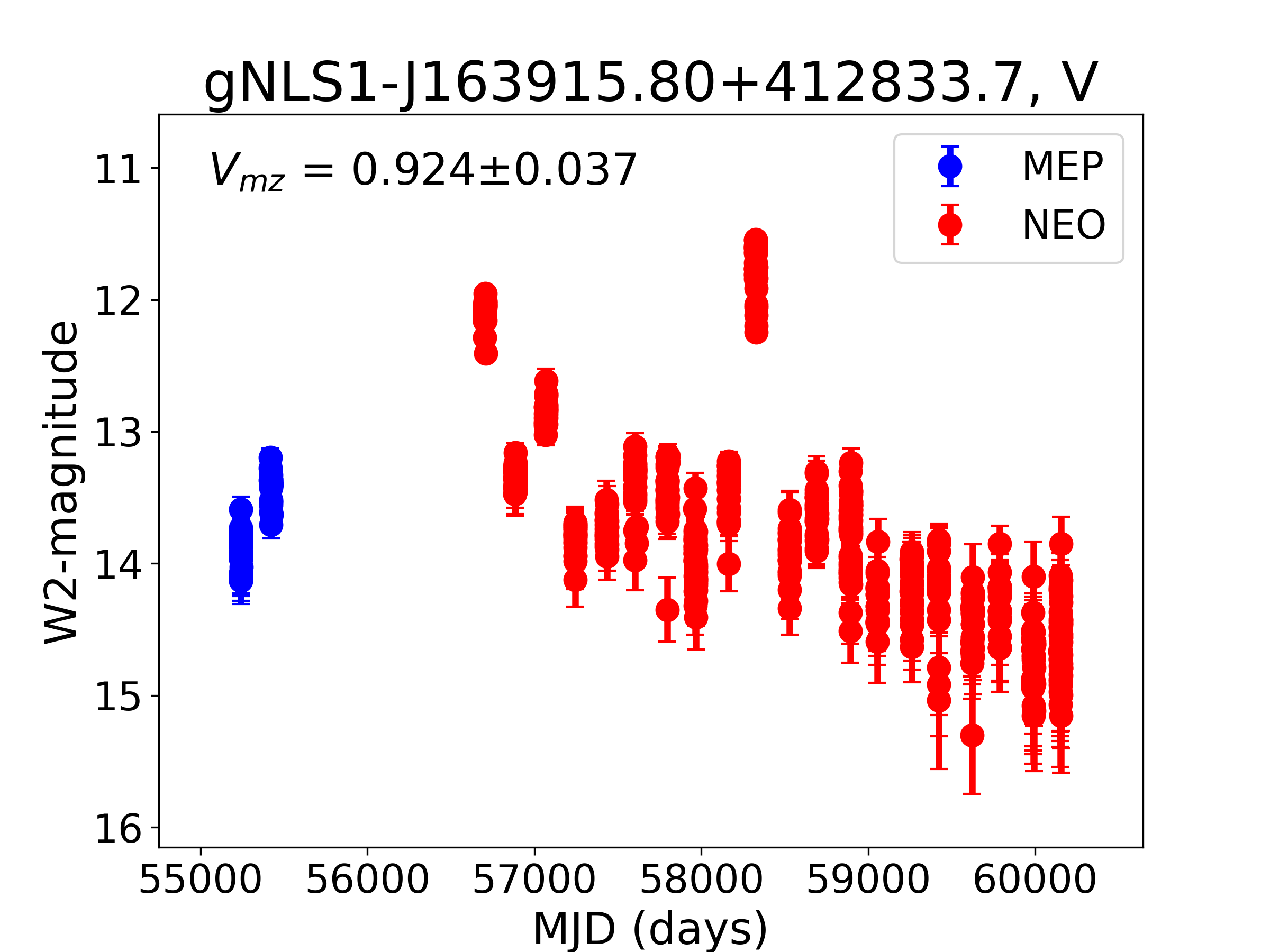}  %left
   \includegraphics[width=0.34\textwidth,height=0.25\textheight,angle=00]{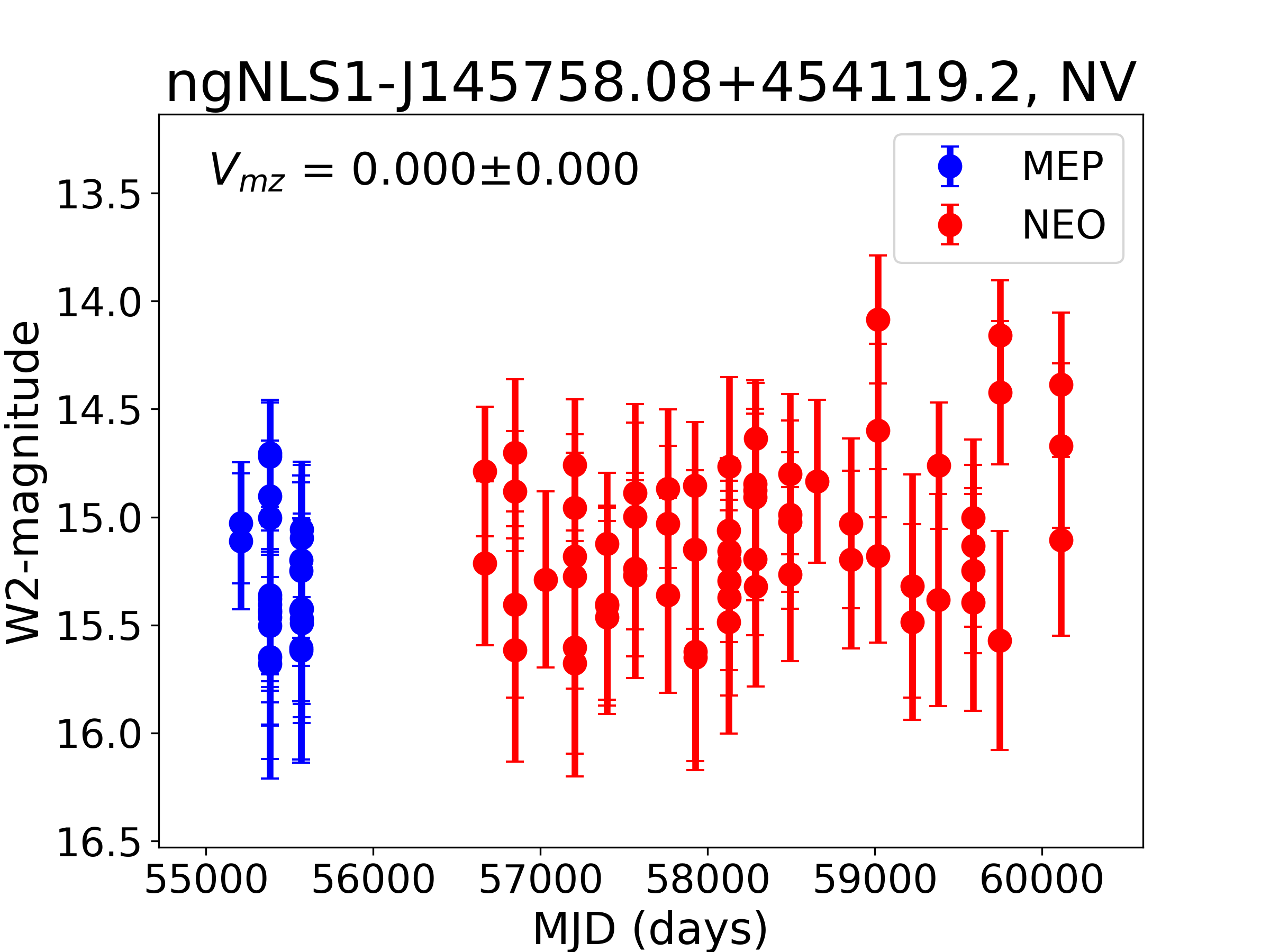} %middle   
   \includegraphics[width=0.34\textwidth,height=0.25\textheight,angle=00]{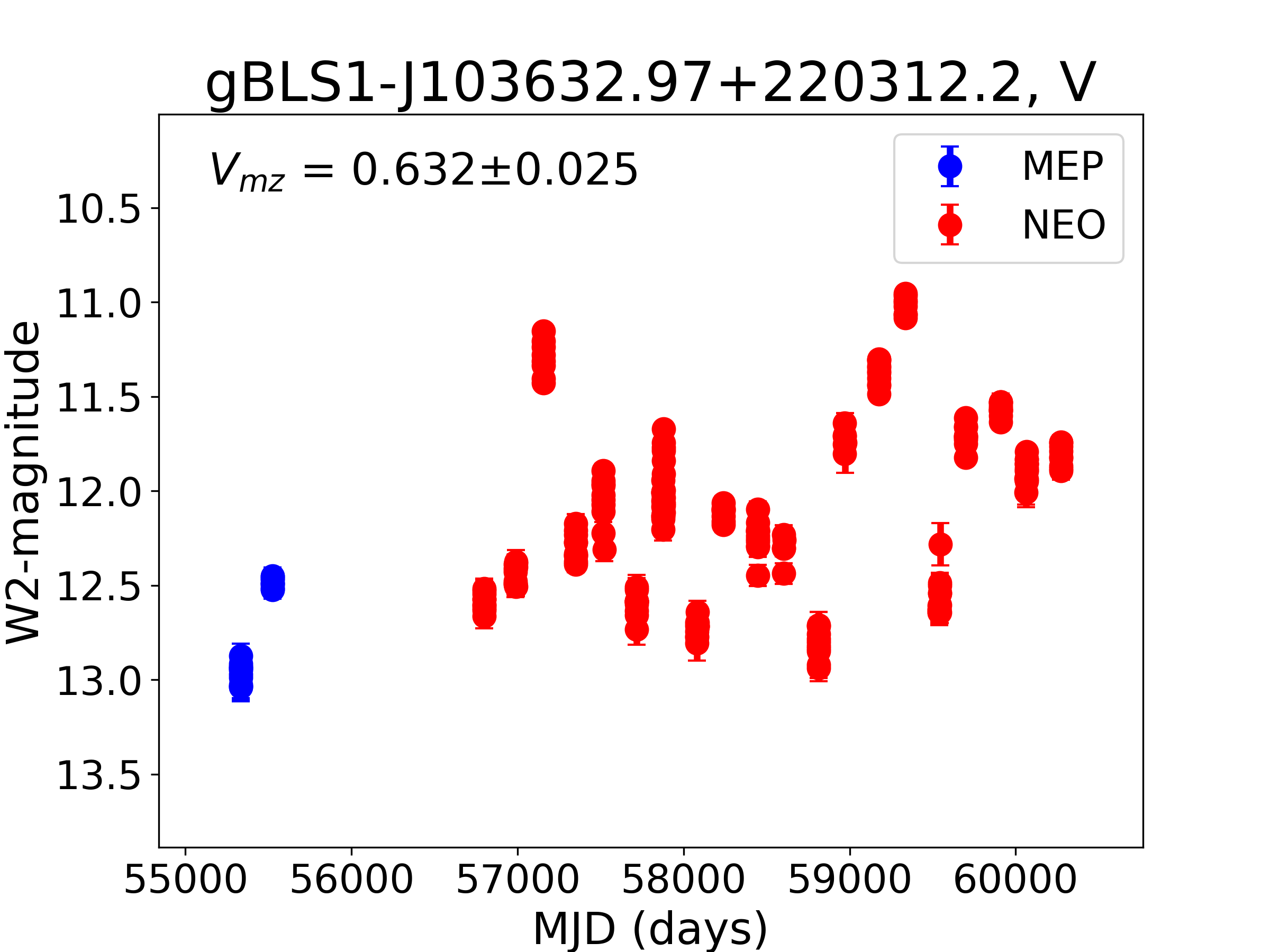}  %right
    \end{minipage}
    \caption{Mid-infrared light curves from WISE for one of the gNLS1-163915.80$+$412833.7 (\emph{left}), ngNLS1-J145758.08$+$454119.2 (\emph{middle}) and gBLS1-J103632.97$+$220312.2 (\emph{right}) from the current sample, showing variability (V), non-variability (NV), and variability (V) on year-like timescale in WISE $\emph {W1}$ and $\emph {W2}$  bands. AllWISE Multi-Epoch Photometry (MEP) and Under Near-Earth Object (NEO) data were presented in blue and red colors, respectively. The redshift corrected intrinsic variability amplitude ($V_{mz}$) is presented in the upper-left corner of each panel.}
    \label{fig: IR_flux_variability}
\end{figure*}

\begin{figure*}
    \begin{minipage}[]{1.0\textwidth}

\includegraphics[width=1.00\textwidth,height=0.25\textheight,angle=00]{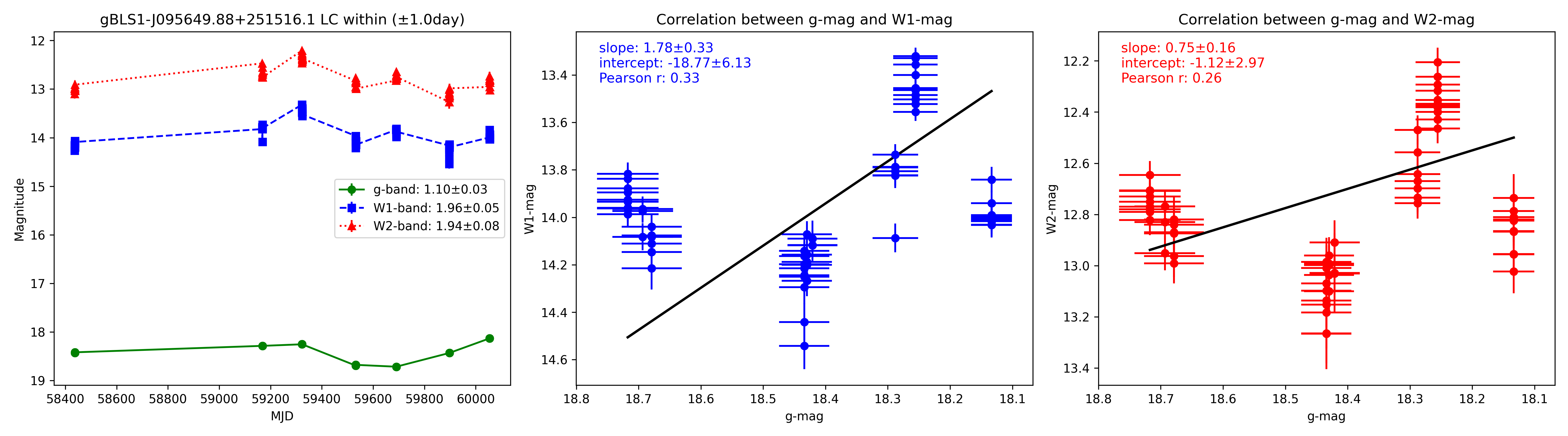}  %left
   \includegraphics[width=1.00\textwidth,height=0.25\textheight,angle=00]{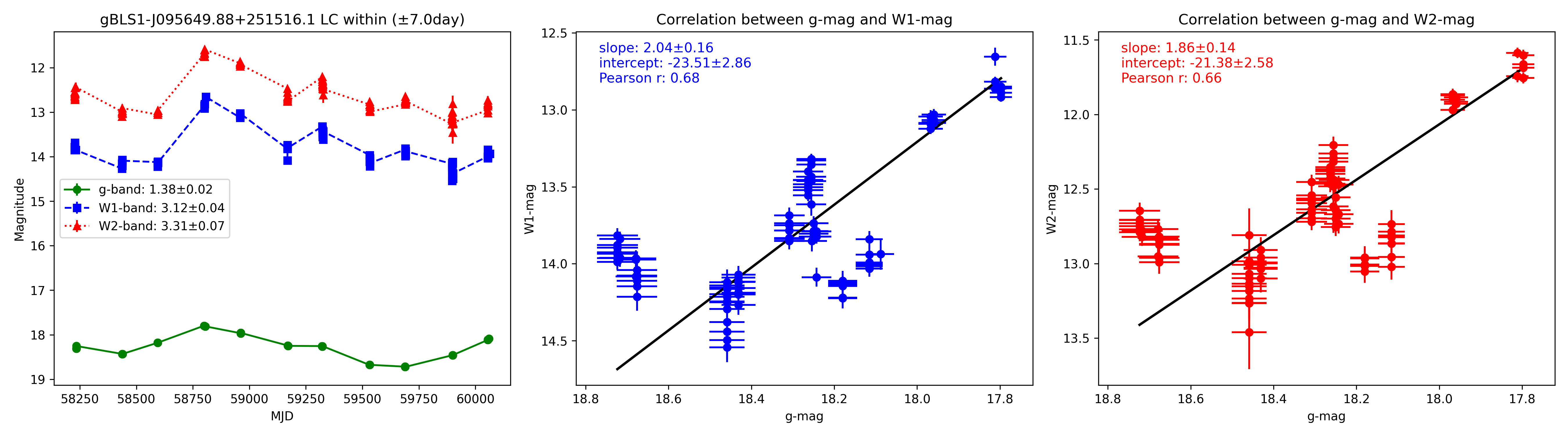}  %left
   \includegraphics[width=1.00\textwidth,height=0.25\textheight,angle=00]{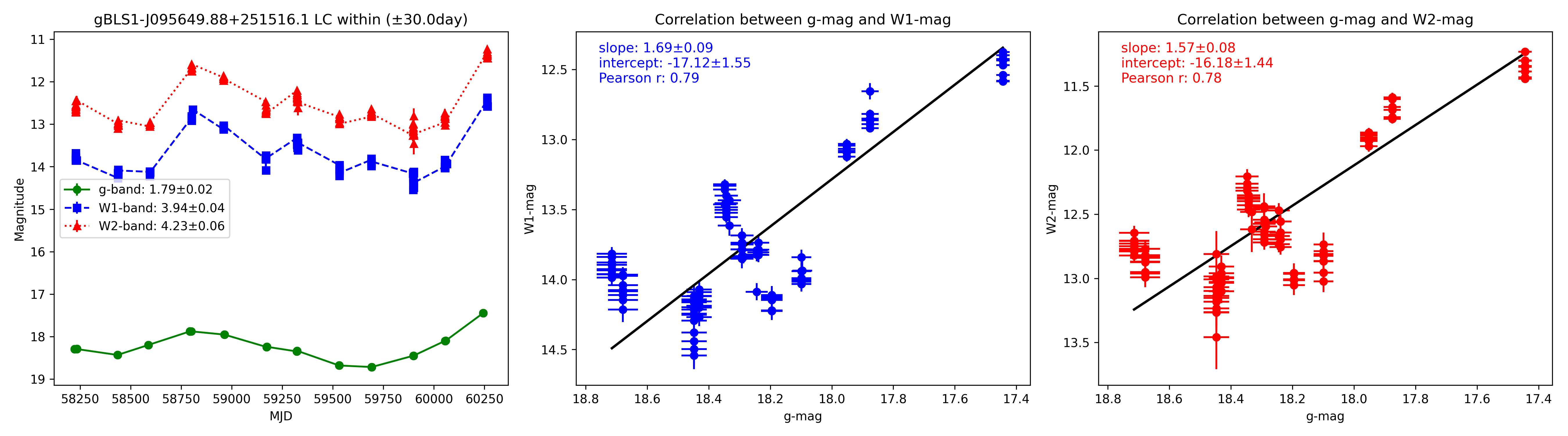}  %left
    \end{minipage}
    \caption{Correlations of MIR $W1$ and $W2$ band light curves with quasi-simultaneous optical $g$-band light curve for gBLS1–J095649.88$+$251516.1. $\emph {Top:}$ $g$-band light curve with MIR $W1$ and $W2$ measurements matched within $\Delta t = \pm1$ day. \emph {Middle:} Optical-MIR correlations for $\Delta t = \pm7$ days. \emph {Bottom:} Optical-MIR correlations for $\Delta t = \pm30$ days. Pearson correlation coefficients (Pearson r) increase with the correlation window: $\rho_r(g$ vs. $W1, W2) = 0.33, 0.26$; $0.68, 0.66$; and $0.79, 0.78$ for $\Delta t = \pm1, \pm7, \pm30$ days, respectively. Black solid lines show orthogonal distance regression fits, accounting for uncertainties in both magnitudes, demonstrating that broader optical sampling windows enhance the observed optical-MIR coupling without altering its intrinsic trend.}
    \label{fig: correlated_optical_and_infrared_var_at_diff_time}
\end{figure*}

\begin{table*}
 %\fontsize{7pt}{7pt}\selectfont
 %\centering
   \caption{Variability characteristics for the samples of gNLS1s, ngNLS1s, and gBLS1s in optical and mid-infrared wavelengths.}  
 \label{ZTF_variability_table}
\begin{tabular}{rcc ccc c| rccc}
    \hline
    \multicolumn{7}{c|}{For the optical wavelength} & \multicolumn{4}{c}{For the mid-infrared wavelength} \\ \hline
    Seyfert galaxies & Duty cycle & Mean & Median & Mean  & Median  & LC$^\star$ & Seyfert galaxies & Mean  & \multicolumn{1}{l}{Error}  & LC$^\star$  \\
    {[no. of sources]} & (in \%) & ($\overline{\psi_{pp}}^{\dag}$) & ($\psi_{\mathrm{pp}}^{\dag}$) &  F$_{\mathrm{var}}$ ($\overline{F_{\mathrm{var}}}$) &  F$_{\mathrm{var}}$ &  of & {[no. of sources]} & $\overline{V_{mz}}$ & err\_$\overline{V_{mz}}$ & of \\
    \hline
gNLS1  [22]  & 68.80  & 1.35$\pm$0.12  & 1.00$\pm$0.10  & 1.11$\pm$0.03 & 0.83$\pm$0.02 & g-band &  gNLS1   [22] & 0.287  & 0.059  & W1-band \\ 
gNLS1  [20]  & 82.67  & 1.25$\pm$0.10  & 0.95$\pm$0.08  & 1.01$\pm$0.02 & 0.85$\pm$0.01 & r-band &  gNLS1   [22] & 0.324  & 0.070  & W2-band \\ 
gNLS1  [16]  & 65.61  & 0.84$\pm$0.08  & 0.63$\pm$0.07  & 0.90$\pm$0.04 & 0.75$\pm$0.04 & i-band &  ngNLS1 [178] & 0.113  & 0.006  & W1-band \\ 
ngNLS1 [171] & 6.42   & 0.86$\pm$0.11  & 0.74$\pm$0.10  & 0.56$\pm$0.07 & 0.53$\pm$0.04 & g-band &  ngNLS1 [178] & 0.026  & 0.004  & W2-band \\ 
ngNLS1 [186] & 5.18   & 0.75$\pm$0.10  & 0.66$\pm$0.09  & 0.42$\pm$0.05 & 0.44$\pm$0.03 & r-band &  gBLS1   [10] & 0.447  & 0.091  & W1-band \\ 
ngNLS1 [147] & 2.45   & 0.56$\pm$0.08  & 0.55$\pm$0.09  & 0.35$\pm$0.08 & 0.34$\pm$0.06 & i-band &  gBLS1   [10] & 0.427  & 0.091  & W2-band \\ 
gBLS1  [10]  & 100    & 1.45$\pm$0.09  & 1.39$\pm$0.06  & 1.46$\pm$0.02 & 1.21$\pm$0.01 & g-band &  ----         & ---    & ---    & -- \\
gBLS1  [10]  & 100    & 1.69$\pm$0.09  & 1.83$\pm$0.06  & 1.81$\pm$0.01 & 1.60$\pm$0.01 & r-band &  ----         & ---    & ---    & -- \\
gBLS1  [08]  & 84.96  & 1.53$\pm$0.11  & 1.45$\pm$0.07  & 1.61$\pm$0.04 & 1.28$\pm$0.03 & i-band &  ----         & ---    & ---    & -- \\
\hline
   
   \multicolumn{11}{l}{$^{\dag}$The mean and median values for the sample considering the light curves belonging to a variable (V) type only. $^\star$Light curve}\\
   \multicolumn{11}{l}{The number of sources used to estimate the statistics are tabulated inside the bracket `[]'.}\\ 
  
 \end{tabular}  
\end{table*}

\begin{table*}
\caption{Variability statistics for the individual gNLS1, ngNLS1, and gBLS1 galaxies in the optical $\emph {g}$, $\emph {r}$, and $\emph {i}$ bands, along with in the mid-infrared $\emph {W1}$ and $\emph {W2}$  bands.}
    \centering
    \begin{tabular}{ccrrcc c c|cc} 
    \hline 
    \multicolumn{8}{c|}{Optical wavelength} & \multicolumn{2}{c}{Mid-infrared wavelength}\\ 
    \hline
      SDSS Name  &Redshift  & Galaxy & \multicolumn{1}{c}{Time}      & Optical&Variability & $\psi_{\mathrm{pp}}$ & $F_{\mathrm{var}}$ & $V_{mz}$  & \multicolumn{1}{c}{Variability}\\
                   & (z) & \multicolumn{1}{c}{type}   & (in days) & \multicolumn{1}{c}{dpts$^{\dagger}$} & \multicolumn{1}{c}{status}    &             &           &      &  \multicolumn{1}{c}{status}  \\              
      \hline 
       \multicolumn{8}{c|}{For the $\emph {g}$-band light curves} & \multicolumn{2}{c}{For the $\emph {W1}$-band light curves}\\ \hline 
    J000132.37$+$211336.2 & 0.439& gNLS1  & 514.781 & 120 &    V  & 0.765$\pm$0.067        & 0.867$\pm$0.022 & 0.140$\pm$0.030 &     V \\ 
    J000545.32$+$063945.4 & 0.884 & ngNLS1 & 2207.987 & 381 &    NV  & 0.660$\pm$0.121  & 0.476$\pm$0.022 & 0.154$\pm$0.016 &     V \\ 
    J094420.44$+$613550.1 & 0.791 & gBLS1  & 2044.211 & 1318 &    V  & 1.043$\pm$0.084      & 0.892$\pm$0.008 & 0.255$\pm$0.038 &     V \\ 
           ----           &---- & ----   &  ----    & ---  & ----      & ---                  & ---             & ----  &    ---    \\
           ----           &---- & ----   &  ----    & ---  & ----      & ---                  & ---             & ----  &    ---    \\                        
      \hline 
       \multicolumn{8}{c|}{For the $\emph {r}$-band light curves} & \multicolumn{2}{c}{For the $\emph {W2}$-band light curves}\\ \hline 
     J000132.37$+$211336.2 & 0.439 & gNLS1  &999.672 & 146 &    V  & 1.090$\pm$0.052            & 1.284$\pm$0.016 & 0.361$\pm$0.034 &     V       \\ 
     J000545.32$+$063945.4 & 0.884 & ngNLS1 & 2198.960 & 506 &    NV  & 0.673$\pm$0.116     &  0.472$\pm$0.018 & 0.066$\pm$0.047 &     NV       \\ 
     J094420.44$+$613550.1 & 0.791 & gBLS1  & 2032.324 & 1475 &    V  & 0.945$\pm$0.078            & 0.804$\pm$0.007 & 0.153$\pm$0.030 &     V       \\ 
            ----           &---- & ----   &  ----    & ---  & ----      & ---                  & ---             & ----  &    ---    \\
            ----           &---- & ----   &  ----    & ---  & ----      & ---                  & ---             & ----  &    ---    \\ 
         \hline 
       \multicolumn{8}{c|}{For the $\emph {i}$-band light curves}\\ \hline 
       J000132.37$+$211336.2 & 0.439 & gNLS1  & 1239.762 & 47 &    V  & 0.852$\pm$0.061            & 1.078$\pm$0.033 & **** &  ****       \\  
       J000545.32$+$063945.4 & 0.884 & ngNLS1 & 1573.788 & 121 &    NV  & 0.639$\pm$0.137      & 0.533$\pm$0.044 & **** &  ****       \\
       J094420.44$+$613550.1 & 0.791 & gBLS1  & 1211.898 & 147 &    NV  & 0.554$\pm$0.104       & 0.389$\pm$0.031 & **** &  ****       \\
              ----           &---- & ----   &  ----    & ---  & ----      & ---                  & ---             & ----  &    ---    \\
              ----           &---- & ----   &  ----    & ---  & ----      & ---                  & ---             & ----  &    ---    \\ 
                                          
      \hline
       \multicolumn{9}{l}{*Data is not available. $^{\dagger}$Number of data points in the optical band light curve. \emph{V:} Variable; \emph{NV:} Non-variable.}\\
      \multicolumn{9}{l}{\emph{Notes:} A portion of this table is presented here to display its form and content, however, the entire table is available in the online}\\ 
      \multicolumn{9}{l}{electronic version.}\\
     \end{tabular}
    \label{tab:OP_IR_variability}
\end{table*}

\subsection{Color variability}
\label{color_variability}
The optical emission observed in AGNs arises from two primary components: the quasi-thermal emission originating from the accretion disk and the nonthermal synchrotron emission from the relativistic jet. Analyzing the color variability of the current sample can provide an effective way to differentiate these components that contribute to the overall flux. Similarly, IR emission, which is considered reprocessed optical/UV radiation~\citep[see, ][]{Storchi-Bergmann1992ApJ...395L..73S, Lu2016MNRAS.458..575L}, offers a unique opportunity to examine the interplay of thermal and nonthermal processes in AGNs. However, to ensure accurate spectral analysis, simultaneous measurements in different bands must be obtained within a 30-minute interval, as described in Sect.~\ref{section_2.0}. This criterion helps mitigate any uncertainties caused by temporal variations and ensures a reliable study of the spectral behavior. Consequently, the color variability analysis in both optical and MIR wavelengths is limited to a subset of sources that meet this stringent timing requirement. The details of the final samples satisfying these conditions are provided at the end of Sect.~\ref{section_2.0}.\par
\begin{figure*}
    \begin{minipage}[]{1.0\textwidth}

  \includegraphics[width=1.0\textwidth,height=0.3\textheight,angle=00]{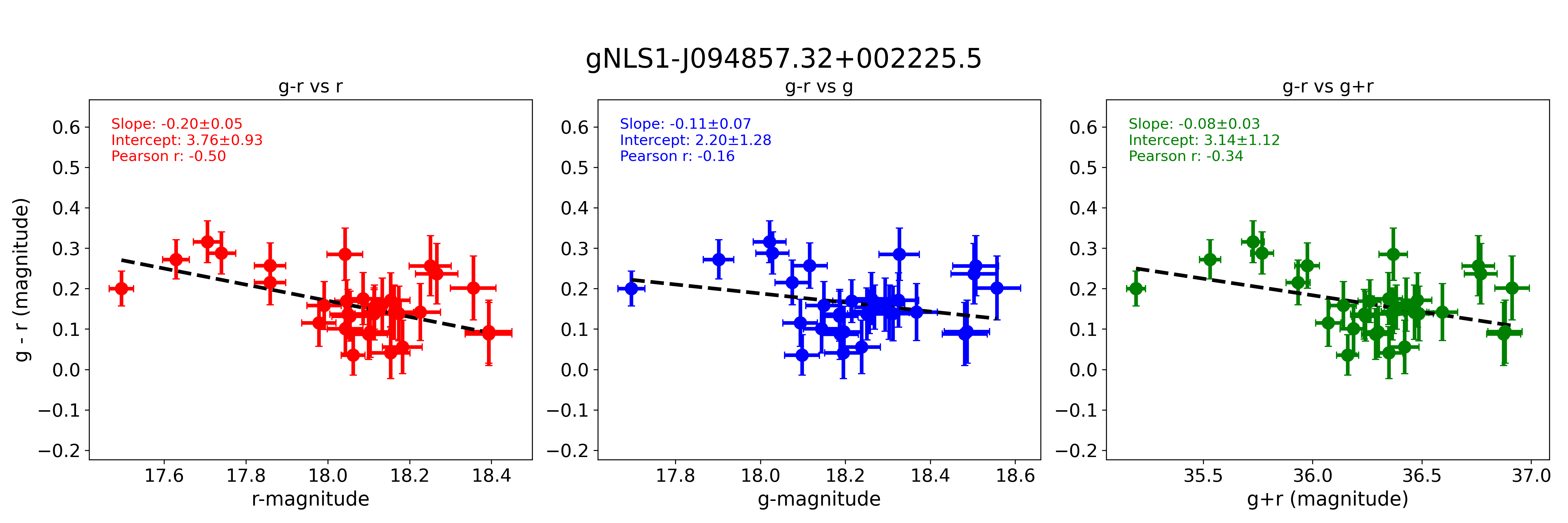}  %left
  \includegraphics[width=1.0\textwidth,height=0.3\textheight,angle=00]{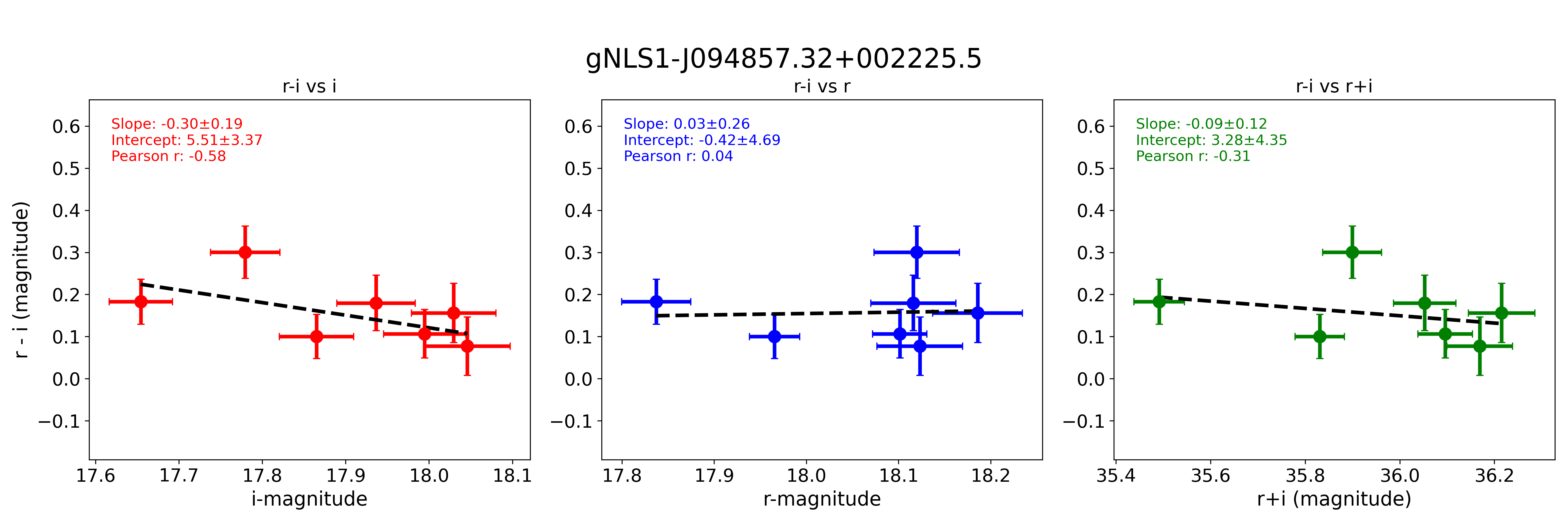}  %middle
   \includegraphics[width=1.0\textwidth,height=0.3\textheight,angle=00]{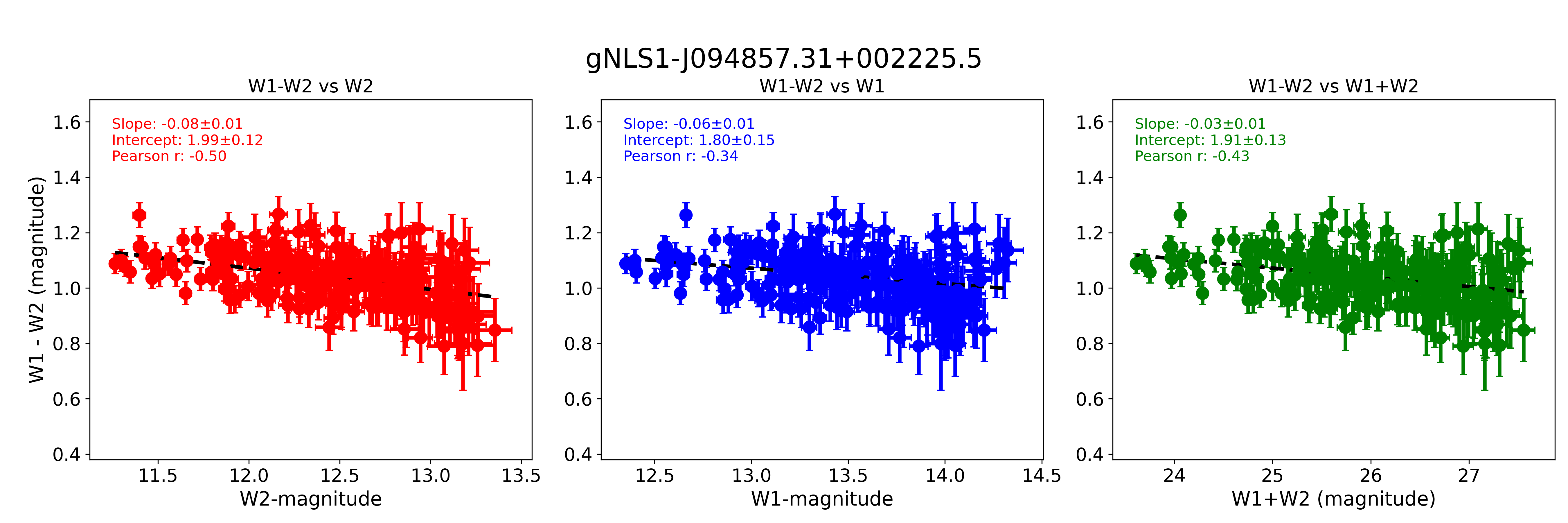} %right  
    \end{minipage}
    \caption{$\emph{Top:}$ Long-term ($\emph {g-r}$) color variation versus $\emph {r}$-magnitude, $\emph {g}$-magnitude and  $\emph {(g+r)}$ (magnitude) plots for representative examples of the gNLS1-J094857.32$+$002225.5 from the current sample.  \emph{Middle:}  Long-term ($\emph {r-i}$) color variation versus $\emph {i}$-magnitude, $\emph {r}$-magnitude and  $\emph {(r+i)}$ (magnitude) plots for the same targets.  \emph{Bottom:} Long-term ($\emph {W1-W2}$) color variation versus $\emph {W2}$-magnitude, $\emph {W1}$-magnitude and $\emph {W1+W2}$ (magnitude) plots for the same targets. All nine panels illustrate a negative trend in the color-magnitude diagrams in both optical and mid-infrared wavelengths. The black dashed line represents the best fit obtained using orthogonal distance regression, accounting for uncertainties in both color and magnitude measurements. The best-fit linear regression parameters, including the slope, intercept, and Pearson correlation coefficient (Pearson r), are displayed in the upper-left corner of each panel.}
    \label{fig: OP_IR_color_variability_gnls1}
\end{figure*}
\begin{figure*}
    \begin{minipage}[]{1.0\textwidth}

\includegraphics[width=1.0\textwidth,height=0.3\textheight,angle=00]{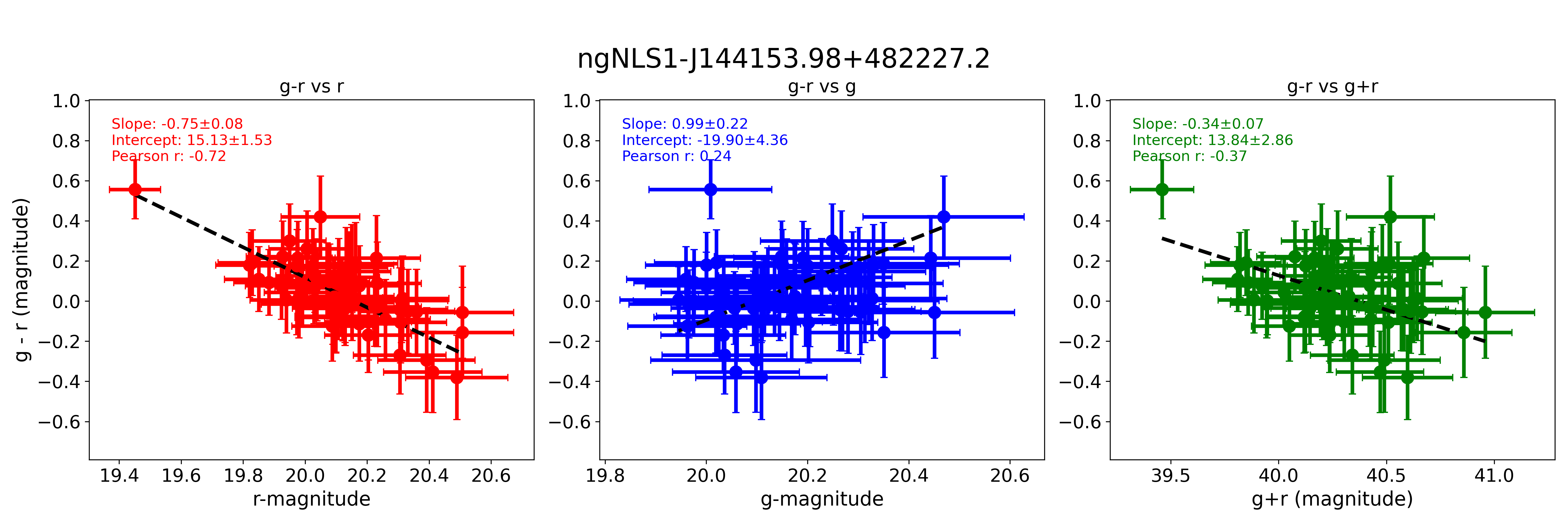}  %left
   \includegraphics[width=1.0\textwidth,height=0.3\textheight,angle=00]{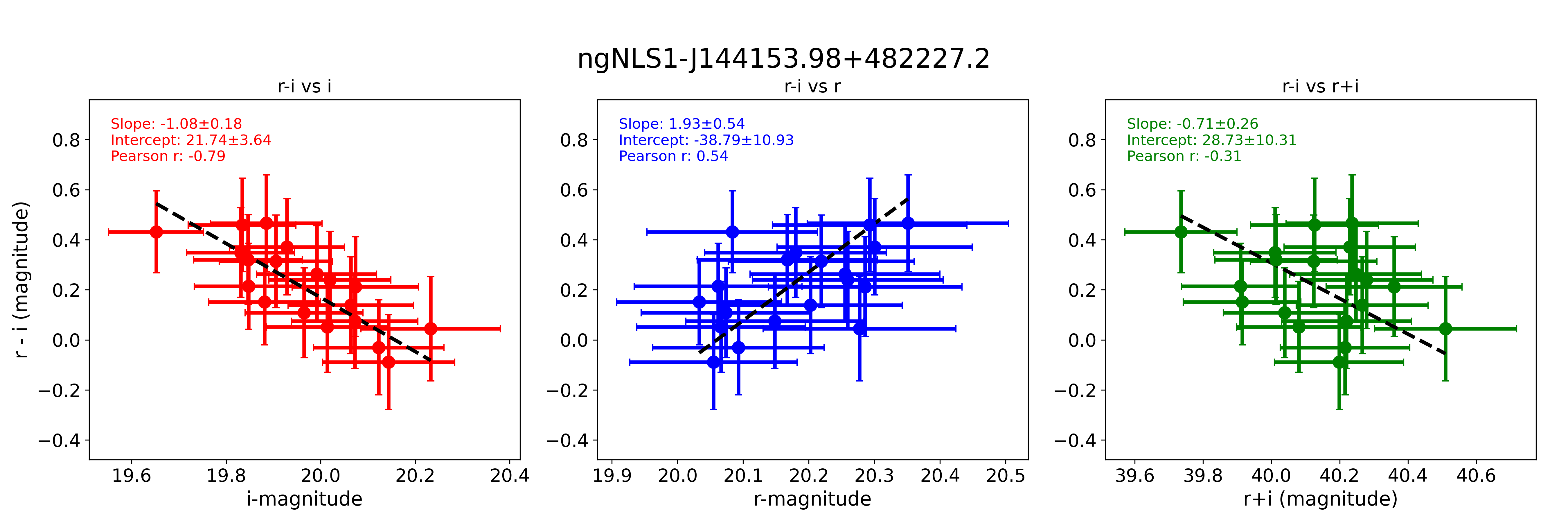}  %middle
   \includegraphics[width=1.0\textwidth,height=0.3\textheight,angle=00]{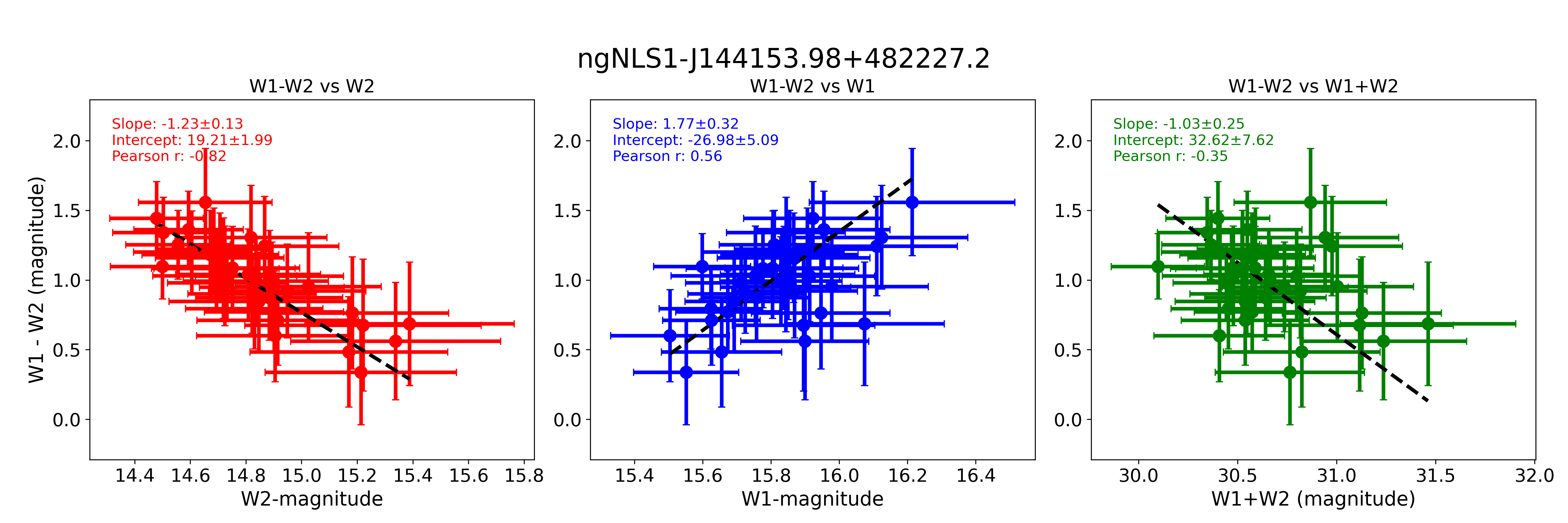}  %right
    \end{minipage}
    \caption{Same as Fig~\ref{fig: OP_IR_color_variability_gnls1} but for ngNLS1-J144153.98$+$482227.2. Here, the middle panel of each subplot shows opposite trends in the color-magnitude diagram as compared to their left and right panels.}
    \label{fig: OP_IR_color_variability_ngnls1}
\end{figure*}
\begin{figure*}
    \begin{minipage}[]{1.0\textwidth}

   \includegraphics[width=1.00\textwidth,height=0.30\textheight,angle=00]{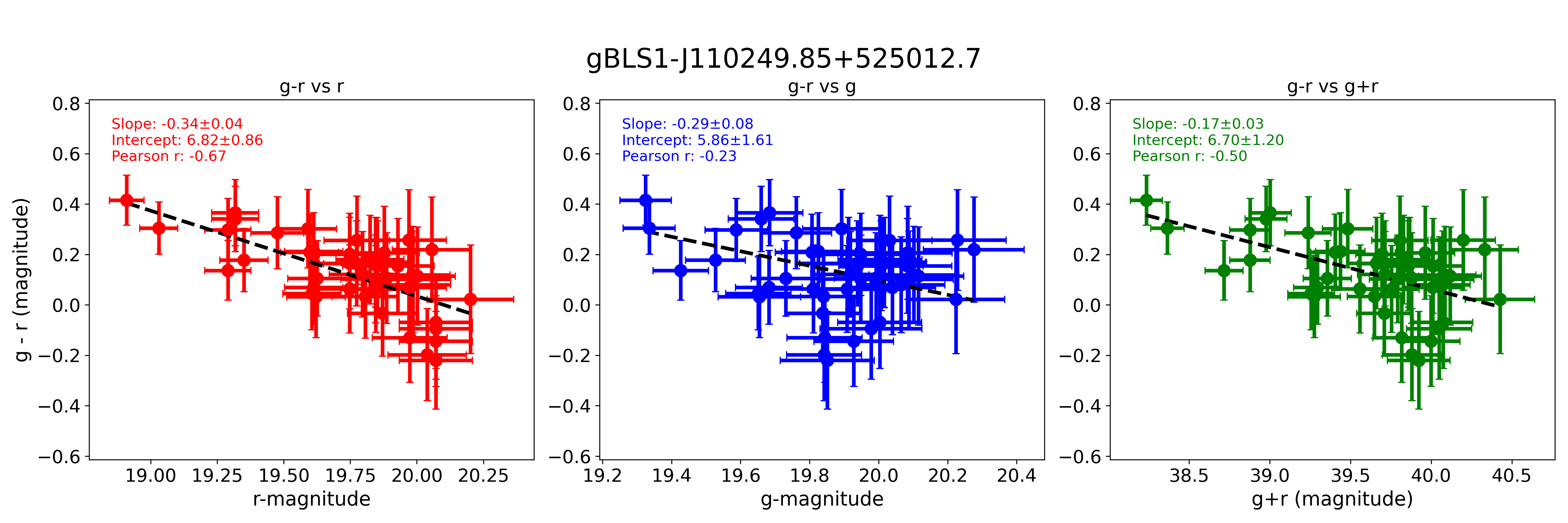}  %left
    \includegraphics[width=1.00\textwidth,height=0.30\textheight,angle=00]{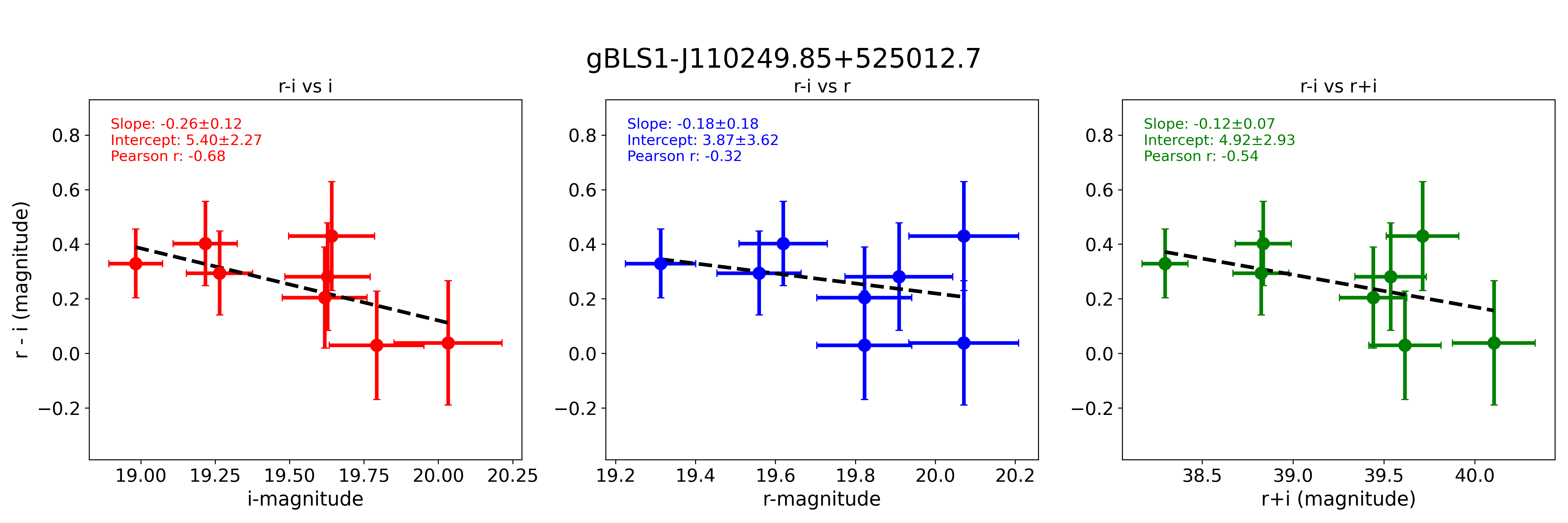}  %middle
\includegraphics[width=1.00\textwidth,height=0.30\textheight,angle=00]{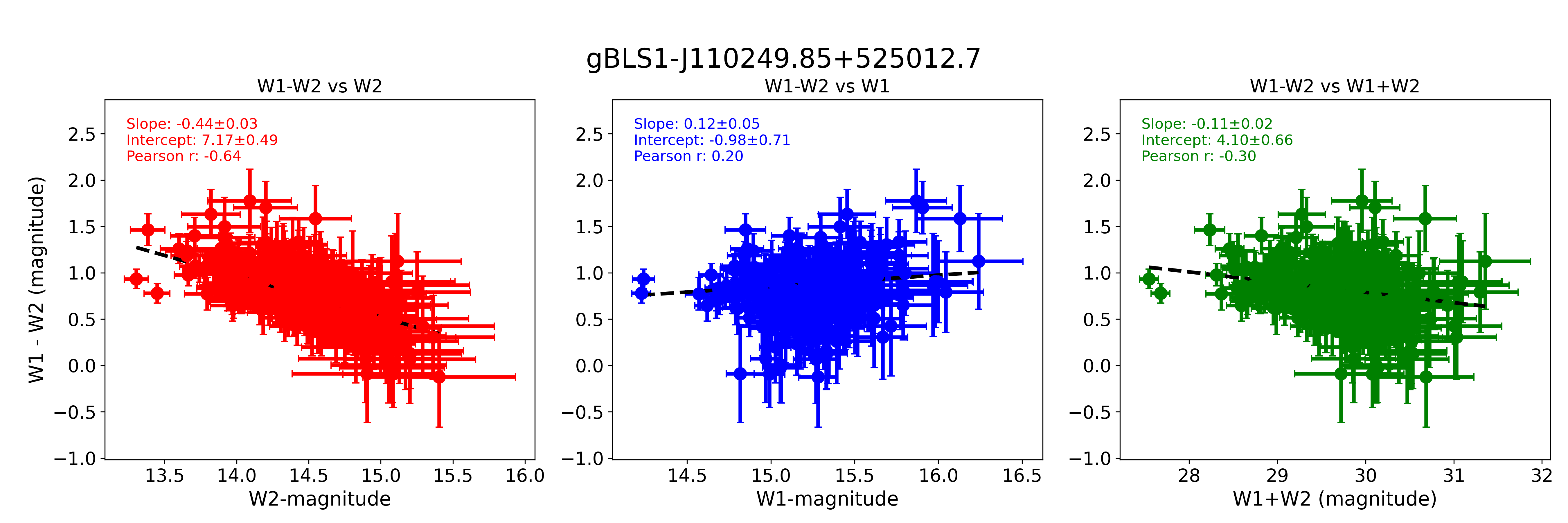}  %right
    \end{minipage}
    \caption{ Same as Fig~\ref{fig: OP_IR_color_variability_gnls1} but for gBLS1-J110249.85$+$555012.7.}
     \label{fig: OP_IR_color_variability_gbls1}
\end{figure*}

For the samples of gNLS1s, ngNLS1s, and gBLS1s mentioned in the last paragraph of Sect.~\ref{section_2.0}, we generated color-magnitude plots, including $\emph {g-r}$ vs. $\emph {r}$, $\emph {r-i}$ vs. $\emph {i}$, and $\emph {W1-W2}$ vs. $\emph {W2}$, using the available quasi-simultaneous data across different optical and MIR bands. Here, the magnitudes of the longer wavelength are chosen on the X-axis, and the difference between the magnitudes of shorter and longer wavelengths is considered for the color to make the color-magnitude analysis consistent across optical and MIR wavelengths. However, during the color-magnitude analysis, we noticed that its trend reverses when we use shorter wavelength (bluer) magnitudes instead of longer wavelength (redder) on the X-axis, especially for ngNLS1s in the MIR wavelength. To account for this, we produced three combinations of color-magnitude diagrams using $\emph  {m1}$ and $\emph {m2}$ magnitudes, that are, $\emph {(m1-m2)}$ vs. $\emph {m2}$, $\emph {(m1-m2)}$ vs. $\emph {m1}$, and $\emph {(m1-m2)}$ vs. $\emph {(m1+m2)}$. Representative examples of these long-term best-fit color-magnitude plots for the gNLS1-J094857.32$+$002225.5, ngNLS1-J144153.98$+$482227.2, and gBLS1-J110249.85$+$525012.7, respectively, are presented in Figs.~\ref{fig: OP_IR_color_variability_gnls1},~\ref{fig: OP_IR_color_variability_ngnls1}, and~\ref{fig: OP_IR_color_variability_gbls1} corresponding to the color-magnitude combinations $\emph {(m1-m2)}$ vs. $\emph {m2}$, $\emph {(m1-m2)}$ vs. $\emph {m1}$, and $\emph {(m1-m2)}$ vs. $\emph {(m1+m2)}$. To examine the relationship between color and magnitude, we employed an ODR fitting on the data points in the color-magnitude plots, accounting for uncertainties in both axes. Furthermore, we calculated $\rho_{r}$ to quantify the correlation between color and magnitude for the gNLS1s, ngNLS1s, and gBLS1s samples across both optical and MIR wavelengths.\par
To qualitatively identify bluer-when-brighter (BWB) or redder-when-brighter (RWB) trends from color-magnitude diagrams, we classified sources based on $\rho_{r}$ values, considering $\rho_{r} \geq 0.5$ indicative of a BWB trend and $\rho_{r} < -0.5$ indicative of an RWB trend. For sources with values of $\rho_{r}$ between $-0.5$ and 0.5, we assumed that there is no significant correlation (termed NOT) between color and magnitude. The number of sources that exhibit trends of RWB, BWB, or NOT, along with their fractions relative to the sample size, is summarized in Table~\ref{tab:OP_IR_color_variability_stat}. The slope, intercept, $\rho_{r}$ and color trends for the combinations of the color-magnitude diagrams $\emph {(m1-m2)}$ vs. $\emph {m2}$, $\emph {(m1-m2)}$ vs. $\emph {m1}$, and $\emph {(m1-m2)}$ vs. $\emph {(m1+m2)}$ of the individual source are tabulated in Table~\ref{tab:OP_IR_color_variability}.


\begin{table*}
\fontsize{6.5pt}{8.5pt}\selectfont % Set font size explicitly
\caption{Distribution of Seyfert galaxies exhibiting the stronger bluer-when-brighter (BWB), redder-when-brighter (RWB), or no significant color-magnitude trends (NOT) in the $\emph {(g-r)}$ color variation versus $\emph {r}$-magnitude plots, $\emph {(r-i)}$ color variation versus $\emph {i}$-magnitude plots, and $\emph {(W1-W2)}$ color variation versus  $\emph {W2}$-magnitude plots. The percentages indicate the proportion of sources in each category relative to the total number of sources in each class.}
\centering
\begin{tabular}{crcrcc cc |c r cc r} 
\hline
\multicolumn{8}{c|}{For optical wavelength} &\multicolumn{4}{c}{For the mid-infrared wavelength} \\ \hline
   Seyfert galaxies & RWB  (\%)   & BWB  (\%)   & NOT  (\%)   &     Seyfert galaxies & RWB  (\%)   & BWB  (\%)   & NOT  (\%)   &    Seyfert galaxies & RWB  (\%)   & BWB  (\%)   & NOT  (\%)   \\
   {[no. of sources]} & \multicolumn{3}{c}{for $\emph {(g-r)}$ color vs. $\emph {(g+r)}-$magnitude} & {[no. of sources]} & \multicolumn{3}{c|}{for $\emph {(r-i)}$ color vs. $\emph {(r+i)}-$magnitude}& {[no. of sources]} & \multicolumn{3}{c}{for $\emph {(W1-W2)}$ color vs. $\emph {(W1+W2)}-$magnitude}     \\
      \hline  
gNLS1 [17]    &  4 (23.53\%) & 5 (29.41\%) & 8 (47.06\%) &  gNLS1 [10]   &  5 (50.00\%) & 1 (10.00\%) &  4 (40.00\%) & gNLS1 [22]    &   12 (54.54\%) &   1 (04.54\%) & 9 (40.91\%) \\
ngNLS1 [128]  & 13 (10.16\%) & 31 (24.22\%) & 84 (65.62\%) &  ngNLS1 [74]  & 36 (48.65\%) & 13 (17.57\%) & 25 (33.78\%) & ngNLS1 [176]  &  49 (27.84\%) &   7 (03.98\%) & 118  (67.04\%) \\
gBLS1 [10]    &  8 (80.00\%) &  0 (0.00\%)& 2  (20.00\%) &  gBLS1 [6]    &  3 (50.00\%) &  1 (0.00\%)&  2                      (50.00\%) & gBLS1 [10]    &    3 (30.00\%) &   3 (10.00\%) & 4 (60.00\%) \\
                 
      \hline 
      \multicolumn{12}{l}{The number of sources used to estimate the statistics are tabulated inside the bracket `[]'.}\\ 
\end{tabular}
\label{tab:OP_IR_color_variability_stat}
\end{table*}

 \section{Results and discussion}
 \label{sec_4.0}
In this study, we systematically investigated flux and color variability for redshift-matched samples of gNLS1s, ngNLS1s, and gBLS1s in optical and mid-infrared wavelengths. For this purpose, we used high-cadence optical light curves in the $\emph {g}$, $\emph {r}$, and $\emph {i}$ bands of ZTF and mid-infrared light curves in the $\emph {W1}$ and $\emph {W2}$ bands from WISE. To analyze flux variability in the optical wavelength, we initially applied the $F^{\eta}$ test using Eq.~\ref{eq.fetest} to the $\emph {g}$, $\emph {r}$, and $\emph {i}$  band light curves to determine the presence or absence of intrinsic variability. Subsequently, $\overline{\psi_{\text{pp}}}$ was calculated exclusively for the variable sources using equation~\ref{eq.Amp_pp}, whereas F$_{\text{var}}$ was calculated for the entire sample, including both variable and non-variable targets, using equation~\ref{eq.Fvar}. The mean and median values of $\psi_{\mathrm{pp}}$ and $F_{\mathrm{var}}$ are tabulated in Table~\ref{ZTF_variability_table}.\par

From Table~\ref{ZTF_variability_table} and Fig.~\ref{fig: OP_flux_variability}, it is evident that gBLS1s exhibit greater variability, as indicated by both $F_{\mathrm{var}}$ and $\psi_{\mathrm{pp}}$, compared to gNLS1s across the $\emph {g}$, $\emph {r}$, and $\emph {i}$  bands, whether considering mean or median values. This finding is consistent with previous studies on smaller and larger samples of gNLS1s and gBLS1s~\citep{Klimek2004ApJ...609...69K, Ai2010ApJ...716L..31A, Ai2013AJ....145...90A, Rakshit2017ApJ...842...96R}. As a class, NLS1s demonstrate lower optical variability amplitudes than BLS1s. This lower variability in NLS1s can be attributed to their higher Eddington ratios (R$_{Edd}$). Since NLS1s accrete at rates significantly higher than BLS1s~\citep{Boroson1992ApJS...80..109B, Peterson2000ApJ...542..161P, Ojha2020ApJ...896...95O}, their optical emission is expected to have a relatively stronger thermal component from the accretion disk compared to the synchrotron emission~\citep[e.g., see][]{Zhou2007ApJ...658L..13Z, Paliya2014ApJ...789..143P}. However, the optical emission from AGNs, dominated by thermal processes, is less variable than the Doppler-boosted synchrotron emission from jets, as a result, the higher contribution of thermal emission in NLS1 due to its higher accretion rate can effectively suppress the amplitude of their optical variability~\citep[e.g., see][]{Ojha2019MNRAS.483.3036O}. However, in contrast to gBLS1s and gNLS1s, ngNLS1s are scarcely variable across the $\emph {g}$, $\emph {r}$, and $\emph {i}$ bands with a mean DC ($\overline{DC}$) of $\sim$ 5\% (see Table~\ref{ZTF_variability_table}, see, e.g., Fig.~\ref{fig: OP_flux_variability}), suggesting that their variability is due to instabilities in the accretion disc.\par 
Based on the observed variability characteristics of gBLS1s, gNLS1s, and ngNLS1s, we interpret the differences in their optical variability amplitudes as arising from the relative contributions of jet and accretion disk emission. The gBLS1s exhibit the highest variability, consistent with a dominant jet component and strong Doppler boosting, which enhances both emission and variability. The gNLS1s, while also hosting relativistic jets, likely have a more substantial contribution from the accretion disk due to their high accretion rates, which acts to dilute the observed variability. Consequently, they exhibit lower variability amplitudes than gBLS1s. In contrast, ngNLS1s, which lack jet emission, show the least variability, consistent with a thermal disk-dominated origin. These results suggest that jets enhance optical variability via relativistic beaming, whereas strong thermal emission from the accretion disk, particularly in high-accretion-rate systems such as NLS1s, suppresses it. The weak variability observed in ngNLS1s is therefore likely driven by intrinsic instabilities within the accretion disk. Although it has been proposed that this disk-driven variability may occur on timescales shorter than those probed by our observations, which alone cannot account for the significantly higher variability seen in jet-dominated systems such as gNLS1s and gBLS1s.

\begin{table}
    \centering
    \caption{Table presenting the slope, intercept, and Pearson correlation coefficient ($\rho_{r}$) for the relationship between $V_{mz}$ and F$_{\mathrm{var}}$ in different optical ($\emph {g}$,  $\emph {r}$,  $\emph {i}$) and infrared ($\emph {W1}$, $\emph {W2}$) bands for the samples of gNLS1s, ngNLS1s, and gBLS1s. The number of sources that satisfy the criteria of optically variable having $V_{mz} \geq 0.1$ is tabulated inside the bracket `[]'. }
    \renewcommand{\arraystretch}{1.1}
    \fontsize{7.0pt}{10.5pt}\selectfont % Set font size explicitly
    \begin{tabular}{|c|c|c|l|r|r|}
        \hline
        {Seyfert } & {OP} & {IR}& {Slope} & {Intercept} & $\rho_{r}$ \\
         {galaxies}& {band}    & {band}    &         &  &  \\
        \hline
        \multirow{6}{*}{gNLS1s} 
        & \multirow{2}{*}{g} & W1 [14] & 6.09$\pm$2.44 & $-$0.22$\pm$0.53 & 0.46 \\
        \cline{3-6} 
        &  & W2 [15] & 6.93$\pm$3.00 & $-$0.29$\pm$0.55 & 0.43 \\
        \cline{2-6}
        & \multirow{2}{*}{r} & W1 [14] & 3.78$\pm$0.96 & 0.15$\pm$0.22 & 0.62 \\  
        \cline{3-6}
        &  & W2 [14] & 3.35$\pm$0.92 & 0.28$\pm$0.19 & 0.50 \\
        \cline{2-6}
        & \multirow{2}{*}{i} & W1 [10] & 4.06$\pm$1.70 & $-$0.06$\pm$0.39 & 0.26 \\  
        \cline{3-6}
        &  & W2 [10] & 3.30$\pm$1.35 & 0.07$\pm$0.33 & 0.15 \\
        \hline
        \multirow{6}{*}{ngNLS1s} 
        & \multirow{2}{*}{g} & W1 [10] & 20.72$\pm$28.07 & $-$2.51$\pm$4.30 & 0.02 \\  
        \cline{3-6}
        &  & W2 [6] & 62.76$\pm$288.79 & $-$7.86$\pm$39.26 & $-$0.23 \\
        \cline{2-6}
        & \multirow{2}{*}{r} & W1 [8] & 11.70$\pm$14.20 & $-$1.28$\pm$2.26 & 0.07 \\  
        \cline{3-6}
        &  & W2 [6] & 89.17$\pm$780.66 & $-$11.70$\pm$107.68 & $-$0.31 \\
        \cline{2-6}
        & \multirow{2}{*}{i} & W1 [3] & 2.22$\pm$0.99 & 0.15$\pm$0.16 & 0.89 \\  
        \cline{3-6}
        &  & W2 [3] & 5.65$\pm$2.81 & $-$0.24$\pm$0.38 & 0.85 \\
        \hline
        \multirow{6}{*}{gBLS1s} 
        & \multirow{2}{*}{g} & W1 [9] & 3.94$\pm$1.19 & $-$0.07$\pm$0.30 & 0.74 \\  
        \cline{3-6}
        &  & W2 [9] & 4.80$\pm$1.89 & $-$0.32$\pm$0.48 & 0.68 \\
        \cline{2-6}
        & \multirow{2}{*}{r} & W1 [9] & 5.64$\pm$1.61 & $-$0.33$\pm$0.40 & 0.72 \\  
        \cline{3-6}
        &  & W2 [9] & 6.77$\pm$2.48 & $-$0.67$\pm$0.62 & 0.68 \\
        \cline{2-6}
        & \multirow{2}{*}{i} & W1 [6] & 1.32$\pm$1.25 & 0.72$\pm$0.50 & 0.18 \\  
        \cline{3-6}
        &  & W2 [6] & 1.50$\pm$1.26 & 0.68$\pm$0.49 & 0.17 \\
        \hline
    \end{tabular}
    \label{tab:correlation_table_Fvar}
\end{table}

\begin{table}
    \centering
    \caption{Same as Fig.~\ref{tab:correlation_table_Fvar} but for the relationship between $V_{mz}$ and $\psi_{\mathrm{pp}}$.}
    \renewcommand{\arraystretch}{1.1}
    \fontsize{7.5pt}{10.5pt}\selectfont % Set font size explicitly
    \begin{tabular}{|c|c|c|r|r|r|}
        \hline
        {Seyfert } & {OP} & {IR}& {Slope} & {Intercept} & $\rho_{r}$ \\
         {galaxies}& {band}    & {band}    &         &  &  \\
        \hline
        \multirow{6}{*}{gNLS1s} & \multirow{2}{*}{g} & W1 [14] & 1.01$\pm$0.40 & 0.58$\pm$0.09 & 0.65 \\
        \cline{3-6}
        &  & W2 [15] & 2.68$\pm$0.80 & 0.12$\pm$0.12 & 0.61 \\
        \cline{2-6}
        & \multirow{2}{*}{r} & W1 [14] & 0.81$\pm$0.33 & 0.62$\pm$0.08 & 0.64 \\
        \cline{3-6}
        &  & W2 [14] & 0.75$\pm$0.24 & 0.62$\pm$0.06 & 0.50 \\
        \cline{2-6}
        & \multirow{2}{*}{i} & W1 [10] & 0.92$\pm$0.35 & 0.26$\pm$0.07 & 0.30 \\
        \cline{3-6}
        &  & W2 [10] & 0.79$\pm$0.25 & 0.28$\pm$0.05 & 0.19 \\
        \hline
        \multirow{6}{*}{ngNLS1s} & \multirow{2}{*}{g} & W1 [10] & $-$0.31$\pm$1.45 & 0.72$\pm$0.23 & 0.21 \\
        \cline{3-6}
        &  & W2 [6] & $-$0.82$\pm$1.35 & 0.76$\pm$0.20 & 0.03 \\
        \cline{2-6}
        & \multirow{2}{*}{r} & W1 [8] & 0.38$\pm$1.70 & 0.48$\pm$0.29 & 0.21 \\
        \cline{3-6}
        &  & W2 [6] & $-$2.62$\pm$2.72 & 0.92$\pm$0.40 & $-$0.17 \\
        \cline{2-6}
        & \multirow{2}{*}{i} & W1 [3] & 5.85$\pm$5.05 & $-$0.39$\pm$0.74 & 0.87 \\
        \cline{3-6}
        &  & W2 [3] & 13.15$\pm$8.64 & $-$1.17$\pm$1.13 & 0.82 \\
        \hline
        \multirow{6}{*}{gBLS1s} & \multirow{2}{*}{g} & W1 [9] & 1.70$\pm$0.54 & 0.35$\pm$0.19 & 0.76 \\
        \cline{3-6}
        &  & W2 [9] & 1.69$\pm$0.63 & 0.35$\pm$0.22 & 0.71 \\
        \cline{2-6}
        & \multirow{2}{*}{r} & W1 [9] & 2.98$\pm$0.69 & 0.06$\pm$0.15 & 0.88 \\
        \cline{3-6}
        &  & W2 [9] & 3.27$\pm$0.90 & $-$0.06$\pm$0.21 & 0.85 \\
        \cline{2-6}
        & \multirow{2}{*}{i} & W1 [6] & $-$0.01$\pm$0.47 & 0.92$\pm$0.24 & 0.27 \\
        \cline{3-6}
        &  & W2 [6] & 0.01$\pm$0.45 & 0.91$\pm$0.22 & 0.26 \\
        \hline
    \end{tabular}
    \label{tab:correlation_table_pp}
\end{table}

\begin{table}
\fontsize{5.5pt}{8.5pt}\selectfont % Set font size explicitly
\caption{Color variability statistics of the individual gNLS1, ngNLS1, and gBLS1 galaxies for the combination of the color-magnitude diagrams $\emph {(m1-m2)}$ vs. $\emph {m2}$, $\emph {m1}$ \& $\emph {m1+m2}$, respectively.}
\centering
\begin{tabular}{ccccccc} 
\hline 
      SDSS Name    & Galaxy                     & Slope & Intercept & $\rho_{r}$ & Color  & Plot  \\
                   & \multicolumn{1}{c}{type}   &       &           &            & trends & for  \\
      \hline

                    & & $-$0.24$\pm$0.09 &   4.89$\pm$ 1.86 & $-$0.53 & RWB & g$-$r vs. r      \\
                    & & $-$0.00$\pm$0.12 &   0.25$\pm$ 2.41 & $-$0.10 & NOT & g$-$r vs. g      \\
                    & & $-$0.07$\pm$0.06 &   2.98$\pm$ 2.24 & $-$0.34 & RWB & g$-$r vs. g$+$r    \\
                    & & $-$0.20$\pm$0.20 &   4.26$\pm$ 3.77 & $-$0.56 & RWB & r$-$i vs. i      \\
J112758.87$+$362028.4 & gNLS1 &  0.05$\pm$ 0.23 &  $-$0.71$\pm$ 4.50 & $-$0.24 & NOT & r$-$i vs. r      \\
                    & & $-$0.05$\pm$ 0.11 &   2.16$\pm$ 4.37 & $-$0.42 & RWB & r$-$i vs. r$+$i    \\
                    & & $-$0.26$\pm$ 0.03 &   4.56$\pm$ 0.48 & $-$0.53 & RWB & W1$-$W2 vs. W2          \\
                    & &  0.03$\pm$ 0.04 &   0.55$\pm$ 0.64 &  0.00 & NOT & W1$-$W2 vs. W1          \\
                    & & $-$0.06$\pm$ 0.02 &   2.88$\pm$ 0.59 & $-$0.30 & RWB & W1$-$W2 vs. W1$+$W2       \\
                    &---- & ---- &  ---- & ---- & ---- & ----       \\
                                        &---- & ---- &  ---- & ---- & ---- & ----       \\
                    & & $-$0.75$\pm$ 0.08 &  15.13$\pm$ 1.53 & $-$0.72 & RWB & g$-$r vs. r      \\
                    & &  0.99$\pm$ 0.22 & $-$19.90$\pm$ 4.36 &  0.24 & NOT & g$-$r vs. g      \\
                    & & $-$0.34$\pm$ 0.07 &  13.84$\pm$ 2.86 & $-$0.37 & RWB & g$-$r vs. g$+$r    \\
                    & & $-$1.08$\pm$ 0.18 &  21.74$\pm$ 3.64 & $-$0.79 & RWB & r$-$i vs. i      \\
J144153.98$+$482227.2 & ngNLS1&  1.93$\pm$ 0.54 & $-$38.79$\pm$10.93 &  0.54 & BWB & r$-$i vs. r      \\
                    & & $-$0.71$\pm$ 0.26 &  28.73$\pm$10.31 & $-$0.31 & RWB & r$-$i vs. r$+$i    \\
                    & & $-$1.23$\pm$ 0.13 &  19.21$\pm$ 1.99 & $-$0.82 & RWB & W1$-$W2 vs. W2          \\
                    & &  1.77$\pm$ 0.32 & $-$26.98$\pm$ 5.09 &  0.56 & BWB & W1$-$W2 vs. W1          \\
                    & & $-$1.03$\pm$ 0.25 &  32.62$\pm$ 7.62 & $-$0.35 & RWB & W1$-$W2 vs. W1$+$W2       \\
                    &---- & ---- &  ---- & ---- & ---- & ----       \\
                                        &---- & ---- &  ---- & ---- & ---- & ----       \\
                    & & $-$0.34$\pm$ 0.04 &   6.82$\pm$ 0.86 & $-$0.67 & RWB & g$-$r vs. r      \\
J110249.85$+$525012.7 & gBLS1 & $-$0.29$\pm$ 0.08 &   5.86$\pm$ 1.61 & $-$0.23 & NOT & g$-$r vs. g      \\
                    & & $-$0.17$\pm$ 0.03 &   6.70$\pm$ 1.20 & $-$0.50 & RWB & g$-$r vs. g$+$r    \\
                    & & $-$0.26$\pm$ 0.12 &   5.40$\pm$ 2.27 & $-$0.68 & RWB & r$-$i vs. i      \\
                    & & $-$0.18$\pm$ 0.18 &   3.87$\pm$ 3.62 & $-$0.32 & NOT & r$-$i vs. r      \\
                    & & $-$0.12$\pm$ 0.07 &   4.92$\pm$ 2.93 & $-$0.54 & RWB & r$-$i vs. r$+$i    \\
                    & & $-$0.44$\pm$ 0.03 &   7.17$\pm$ 0.49 & $-$0.64 & RWB & W1$-$W2 vs. W2          \\
                    & &  0.12$\pm$ 0.05 &  $-$0.98$\pm$ 0.71 &  0.20 & NOT & W1$-$W2 vs. W1          \\
                    & & $-$0.11$\pm$ 0.02 &   4.10$\pm$ 0.66 & $-$0.30 & NOT & W1$-$W2 vs. W1$+$W2       \\
                    &---- & ---- &  ---- & ---- & ---- & ----       \\
                                        &---- & ---- &  ---- & ---- & ---- & ----       \\
\hline
\multicolumn{7}{l}{\emph{Notes:} A portion of this table is presented here to display its form and content, however, the entire table}\\ 
      \multicolumn{7}{l}{is available in the online electronic version.}\\
\end{tabular}
\label{tab:OP_IR_color_variability}
\end{table}

In contrast to the accretion disc based variability in ngNLS1s, the origin of resulted flux variability of gNLS1s and gBLS1s could either be due to accretion disc instabilities~\citep[e.g., ][]{Wiita1991sepa.conf..557W,Chakrabarti1993ApJ...411..602C, Mangalam1993ApJ...406..420M} or fluctuations emerging within the jet~\citep[e.g., ][]{Marscher1985ApJ...298..114M, Wagner1995ARA&A..33..163W, Marscher2014ApJ...780...87M}. Therefore, we have adopted the variability amplitude cut of $\geq$ 0.4 mags suggested by~\citet{Bauer2009ApJ...705...46B} to distinguish between jet-based variability and accretion disc instability-based variability. Thus, we first checked $\psi_{\text{pp}}$ for the variable targets in the samples of gNLS1s and gBLS1s, and found that 15 out of 17 variable gNLS1s ($\sim$88\%) and 10 out of 10 variable gBLS1s (100\%) exhibit $\psi_{\text{pp}} \geq$ 0.4 mags for the $\emph {r}$-band  light-curves. Among the gNLS1s, 89\% and 75\% of the sample are found to be variable in the $\emph{g}$ and $\emph{i}$ bands, respectively. In contrast, all sources (100\%) in the gBLS1 sample exhibit variability in both the $\emph{g}$ and $\emph{i}$ bands. From this analysis, it appears that gBLS1s tend to be more variable compared to gNLS1s, which could be due to the larger observed contribution of jets more closely aligned towards the observer line of sight in gBLS1s~\citep[e.g.][]{Safna2020MNRAS.498.3578S}. A comparatively small percentage of the variable targets with $\psi_{\text{pp}} \geq$ 0.4 mags in the sample of gNLS1s could also be due to a difference in their accretion rates as discussed above (see 2$^{nd}$ para). On the other hand, a similar kind of variability nature resulted from the samples of gNLS1s, ngNLS1s and gBLS1s in the mid-infrared $\emph {W1}$ and $\emph {W2}$ bands (see Table~\ref{ZTF_variability_table}, Fig.~\ref{fig: IR_flux_variability}), supporting a reprocessing of optical/UV emission from the central engine of Seyfert galaxies by the dusty torus to MIR wavelength. Furthermore, to strengthen the reprocessing scenario found here, we have checked the correlation of $V_{mz}$ with $F_{\mathrm{var}}$ and $\psi_{\mathrm{pp}}$ for optically variable targets having $V_{mz} \geq$ 0.1 in the mid-infrared wavelength for the samples of gNLS1s, ngNLS1s, and gBLS1s.\par 
A strong and statistically significant correlation ($\rho_{r} \geq 0.7$) is found between the MIR variability amplitude ($V_{mz}$) and both the optical $F_{\mathrm{var}}$ and $\psi_{\mathrm{pp}}$ in the sample of gBLS1s, particularly for the $\emph{g}$ and $\emph{r}$-band light curves with the $\emph{W1}$ and $\emph{W2}$-band light curves (see Fig.~\ref{fig: correlated_optical_and_infrared_var}; Tables~\ref{tab:correlation_table_Fvar},~\ref{tab:correlation_table_pp}). In contrast, while the gNLS1 sample also exhibits statistically significant correlations ($\rho_{r} \gtrsim  0.5$), these are consistently weaker than those found for gBLS1s in the same bands. This trend suggests that MIR variability in both gBLS1s and gNLS1s is primarily driven by instabilities in the accretion disk. However, the comparatively weaker correlations in gNLS1s may be due to additional contamination in their optical variability amplitudes, likely arising from stronger host galaxy contributions. Moreover, this interpretation supports the view that MIR variability in gBLS1s is more strongly disk-dominated, whereas in gNLS1s, the combined influence of jets and host contamination may dilute the disk-driven variability signature in the optical bands. Significant correlations are primarily seen in the $\emph{g}$ and $\emph{r}$ bands, while correlations involving the $\emph{i}$-band light curves are generally weak or statistically insignificant. This could be attributed to the relatively sparse sampling in the $\emph{i}$ band, making it difficult to draw definitive conclusions from those data. No significant correlations are found for the ngNLS1 sample in any optical band, except for a notable trend in the $\emph{i}$ band. However, given the small sample size (only 8 and 6 sources for the $\emph{W1}$ and $\emph{W2}$ bands, respectively), along with potentially strong host galaxy contamination, the lack of significant correlations in ngNLS1s may result from these limitations rather than the absence of an intrinsic physical connection between optical and MIR variability.\par

Given that the optical $g$, $r$, and $i$ bands are sampled more finely than the MIR $\emph{W1}$ and $\emph{W2}$ bands for our sample of Seyfert galaxies, we investigated whether variability at optical and infrared wavelengths is sensitive to differences in cadence. To do this, we selected sources that were optically variable and had $V_{mz} \ge 0.1$ in the MIR, and constructed quasi-simultaneous optical-MIR pairs using windows of $\Delta t = \{\pm1, \pm7, \pm30\}$ days centered on each MIR epoch. Within each window, multiple optical measurements were replaced by their median, with uncertainties propagated via the rms of reported errors. For each $\Delta t$, we recomputed the fractional variability amplitude, $F_{\rm var}$, in all bands, and quantified the optical-MIR coupling using the newly constructed optical and MIR light curves (e.g., see Fig.~\ref{fig: correlated_optical_and_infrared_var_at_diff_time}), applying ODR that accounts for errors on both axes, along with $\rho_r$. Increasing $\Delta t$ modestly raises $F_{\rm var}$ across all bands, as expected when integrating variability over longer timescales, but does not systematically weaken optical-MIR correlations where they are already present at finer matching. Sources with sparse pairs within $\pm1$d often yield unconstrained results, which become well defined at $\pm7{-}30$d, whereas sources with good overlap show stable slopes ($\sim1{-}2$) and high values $\rho_{r}$ ($\sim0.6{-}0.95$) across all windows. We conclude that (i) $F_{\rm var}$ primarily probes variability on the week–month scale, and (ii) optical and MIR variations occur on comparable timescales. Resampling optical light curves to MIR-like cadence by widening the matching window does not weaken the optical--MIR correlation; if anything, it clarifies it in sparsely sampled cases, indicating that both bands trace the same underlying variability and that the observed cross-band agreement is not a sampling artifact.

The relatively weak variability (duty cycle $\sim$5\%) and the lack of significant correlation between the optical and infrared variability amplitudes observed in ngNLS1s may be attributed to their characteristically weak yet rapid variability, manifesting over short (intranight) timescales. Such rapid fluctuations are not well captured by large-scale time-domain surveys such as ZTF and WISE, which typically operate with cadences of a day or longer. This interpretation aligns well with the findings of the Super-Eddington Accreting Massive Black Holes (SEAMBH) reverberation mapping (RM) project~\citep[e.g., see][]{Du2018ApJ...856....6D, Dunlop2003MNRAS.340.1095D}, which systematically monitored a sample dominated by ngNLS1s. The SEAMBH project demonstrated that H$\beta$ reverberation lags in these high accretion rate AGNs are significantly shorter by factors of $\sim$ 2-6 than those predicted by the canonical $R_{H\beta}-L_{5100}$ relation. These shortened lags reflect the compact nature of the broad-line region in SEAMBHs and underscore the importance of high-cadence observations for capturing their rapid variability. Thus, the weak variability signatures observed in the present study probably reflect a cadence-induced bias rather than an absence of intrinsic activity.\par
On the other hand, as discussed in Sect.~\ref{color_variability}, studying the color variability of AGNs can provide an effective way to disentangle between quasi-thermal emission from the accretion disc and nonthermal synchrotron emission from the jet, which contributes to the overall observed emission. Therefore, for our samples of gNLS1s, ngNLS1s, and gBLS1s, we generated color-magnitude ($\emph {g-r}$ vs. $\emph {r}$ \& $\emph {r-i}$ vs. $\emph {i}$) diagrams in the optical regime and $\emph {W1-W2}$ vs. $\emph {W2}$ in the MIR regime using their quasi-simultaneous data observed in respective bands. However, during the color-magnitude analysis, we noticed that its trend reverses when we use shorter wavelength (bluer) magnitudes in place of longer wavelength (redder) on the X-axis, especially for ngNLS1s in the MIR wavelength. To account for this, we produced three combinations of color-magnitude diagrams using $\emph  {m1}$ and $\emph {m2}$ magnitudes, that are, $\emph {(m1-m2)}$ vs. $\emph {m2}$, $\emph {(m1-m2)}$ vs. $\emph {m1}$, and $\emph {(m1-m2)}$ vs. $\emph {(m1+m2)}$. In Fig.~\ref{fig: OP_IR_color_variability_gnls1}, \emph{Top:} long-term ($\emph {g-r}$) color variation versus $\emph {r}$-magnitude, $\emph {g}$-magnitude and $\emph {(g+r)}$ (magnitude) plots for representative example of the gNLS1-J094857.32$+$002225.5 from the current sample. \emph{Middle:}  Same as top but for the ($\emph {r-i}$) color vs. $\emph {i}$, $\emph {r}$ and $\emph {(r+i)}$ magnitudes for the same target.  \emph{Bottom:} Same as top, but for the ($\emph {W1-W2}$) color vs. $\emph {W2}$, $\emph {W1}$ and $\emph {(W1+W2)}$ magnitudes for the same target. All nine panels illustrate negative trends in the color-magnitude diagrams in both optical and mid-infrared wavelengths. The black dashed line represents the best-fit curve obtained using ODR, which accounts for uncertainties in both color and magnitude measurements. The best-fit linear regression parameters, including the slope, intercept, and Pearson's correlation coefficient (Pearson r), are displayed in the upper left corner of each panel. A very similar color-magnitude diagrams are presented for ngNLS1-J144153.98$+$482227.2 and gBLS1-J110249.85$+$525012.7 in the Figs.~\ref{fig: OP_IR_color_variability_ngnls1}~\&~\ref{fig: OP_IR_color_variability_gbls1}, respectively.\par 
Since a genuine correlation should not depend on the specific choice of magnitude in a color–magnitude diagram, the observed reversal of trends in cases where the disk contribution dominates led us to base our conclusions primarily on the $\emph{(m1-m2)}$ color versus $\emph{(m1+m2)}$ magnitude diagrams. We further noted that, for this method, the strength of the correlation is consistently weaker compared to the other two combinations, even in cases of true correlations (i.e., those without trend reversals). Therefore, we adopt a more relaxed threshold, requiring $|\rho_{r}| \geq 0.3$, to define strong and reliable correlations. Under this criterion, $\rho_{r} \geq 0.3$ corresponds to a BWB trend, while $\rho_{r} \leq -0.3$ indicates a RWB trend.\par
Applying this framework to the $\emph{(g-r)}$ vs. $\emph{(g+r)}$ diagrams, we find that $\sim24\%$ (4/17) and $\sim29\%$ (5/17) of gNLS1s exhibit RWB and BWB behavior, respectively (see Table~\ref{tab:OP_IR_color_variability_stat}). For ngNLS1s, the corresponding fractions are $\sim10\%$ (13/128) and $\sim24\%$ (31/128), respectively. In the $\emph{(r-i)}$ vs. $\emph{(r+i)}$ diagrams, we find $\sim50\%$ (5/10) and $\sim10\%$ (1/10) of gNLS1s exhibiting RWB and BWB behavior, respectively, while the ngNLS1 sample yields $\sim49\%$ (36/74) RWB and $\sim17\%$ (13/74) BWB trends. For gBLS1s, $\sim80\%$ (8/10) exhibit the RWB trend in $\emph{(g-r)}$ vs. $\emph{(g+r)}$, and $\sim50\%$ (3/6) exhibit the same trend in $\emph{(r-i)}$ vs. $\emph{(r+i)}$, with the exception of J094420.44$+$613550.1, which shows a BWB trend in the latter diagram. We also note that a significant fraction of sources show no strong correlations under the adopted criterion. Specifically, 8 out of 17 gNLS1s ($\sim47\%$), 84 out of 128 ngNLS1s ($\sim66\%$), and 2 out of 10 gBLS1s ($\sim20\%$) do not exhibit strong correlations between $\emph{(g-r)}$ color and $\emph{(g+r)}$ magnitude. For the $\emph{(r-i)}$ vs. $\emph{(r+i)}$ diagrams, the corresponding fractions are 4 out of 10 gNLS1s ($\sim40\%$), 25 out of 74 ngNLS1s ($\sim34\%$), and 2 out of 6 gBLS1s ($\sim33\%$) (see Table~\ref{tab:OP_IR_color_variability_stat}).\par 
On the other hand, similar to the optical wavelength, $\sim$ 55\% (15/22), $\sim$ 28\% (49/176), and $\sim$ 30\% (3/10) of gNLS1s, ngNLS1s, and gBLS1s, respectively, showed the RWB trend in color magnitude diagram of $\emph {(W1-W2)}$ vs. $\emph {(W1+W2)}$ (see Table~\ref{tab:OP_IR_color_variability_stat}). Contrary to the RWB trend exhibited by most sources, a smaller percentage of $\sim$ 4\% (1/22), $\sim$ 4\% (7/176), and $\sim$ 30\% (3/10) of gNLS1s, ngNLS1s, and gBLS1s, respectively, showed the BWB trend (see Table~\ref{tab:OP_IR_color_variability_stat}). The remaining $\sim$ 41\% (9/22), $\sim$ 67\% (118/176), and $\sim$ 40\% (4/10) of gNLS1s, ngNLS1s, and gBLS1s, respectively, didn't exhibit a strong correlation between $\emph {(W1-W2)}$ color and $\emph {(W1+W2)}$ magnitude (see Table~\ref{tab:OP_IR_color_variability_stat}).\par
Based on the color–magnitude analysis \emph{(m1-m2)} vs. \emph{(m1+m2)} for gNLS1s, ngNLS1s, and gBLS1s across optical to MIR wavelengths, we find the following trends: in the shorter wavelength regime, BWB behavior dominates over RWB in the color–magnitude diagrams (e.g., \emph{(g-r)} vs. \emph{(g+r)}). In contrast, in the longer wavelength regimes, RWB behavior dominates over BWB (e.g., \emph{(r-i)} vs. \emph{(r+i)} and \emph{(W1-W2)} vs. \emph{(W1+W2)}), except for gBLS1s, where RWB behavior is dominant from the optical through the MIR.\par
The so-called BWB and RWB color trends from the AGN in the optical wavelength attribute to domination of nonthermal emission from the jet over thermal emission from the accretion disc, and vice-versa, respectively~\citep[e.g., see][]{Malkan1983ApJ...264L...1M, Pian1999ApJ...521..112P, Rani2010MNRAS.404.1992R, Sakata2010ApJ...711..461S, Ikejiri2011PASJ...63..639I, Mao2016Ap&SS.361..345M, Negi2022MNRAS.510.1791N}. The BWB trend that predominantly appeared in the BL Lacs class of AGN can be explained with the shock-in-jet model~\citep[see ][]{2011MmSAI..82..104T} where electrons are accelerated to higher energies at the front of the shock, and since high-energy electrons lose their energy faster compared to low-energy electrons by radiative cooling process, thus making the high-energy band (in the current work $\emph {g}$-band compared to $\emph {r}$-band and $\emph {r}$-band compared to $\emph {i}$-band) more variable~\citep{Kirk1998A&A...333..452K, Mastichiadis2002PASA...19..138M}. It could also be explained by an energy injection of fresh electrons due to internal shocks in the relativistic shell into the emitting regions, leading to an increase in the number of high energy electrons and eventually shifting $\nu_{peak}$ to higher energy~\citep[e.g., ][]{Spada2001MNRAS.325.1559S, Zhang2002ApJ...572..762Z, Fiorucci2004A&A...419...25F}. Apart from the energy injection scenario, the BWB trend can also be attributed to variations in the beaming factor, $\delta$, of the emitting region. Changes in $\delta$ not only result in an apparent increase in observed flux, following the relation $_{\nu}f_{\nu} \propto \delta^{4}$, but also cause a shift in the observed frequency, $\nu_{obs} \propto \delta$~\citep[e.g., ][]{Villata2004A&A...421..103V, Papadakis2007A&A...470..857P, Larionov2010A&A...510A..93L}.\par
On the other hand, the RWB trend that appeared in the Seyfert galaxies in the current work, especially towards longer wavelengths, can be explained by the domination of the disc's fluctuations over the jet's fluctuations. Since long-term optical variability in Seyfert galaxies is exhibited by a mix of disc and jet emissions, if the accretion disc's fluctuations are primarily responsible for the variability but jet contributes more in blue/UV, then any brightening of the disc may increase the domination of redder thermal emission, leading to RWB trend~\citep[e.g., see][]{Malkan1983ApJ...264L...1M, Pian1999ApJ...521..112P, Sakata2010ApJ...711..461S}. The RWB trend observed in Seyfert galaxies and FSRQs can also be accommodated in the shock-in-jet model. When a shock propagates through a jet and interacts with regions of higher electron density, stronger magnetic fields, or turbulence, the emission from that region intensifies. Initially, higher-energy radiation emerges first, followed by a delayed peak at lower frequencies~\citep[e.g.][]{Valtaoja2002PASA...19..117V}. In the later stages of a flare, the flux enhancement in the redder bands may become more pronounced as the contribution from shorter wavelengths stabilizes or declines, leading to an overall RWB trend. This effect is particularly relevant in Seyfert galaxies, where the jet emission is weaker, and contributions from the disc and dust are more significant. In FSRQs, the strong thermal emission from the accretion disc, which peaks in UV, plays a crucial role. As the synchrotron emission from the jet increases, typically peaks at lower frequencies, it can shift the overall spectral energy distribution to redder colors, reinforcing the RWB trend~\citep[e.g., see][]{Gu2006A&A...450...39G}.\par
Thus, based on the long-term optical color variability observed in the current samples of gNLS1s, ngNLS1s, and gBLS1s, the dominant mechanism driving their long-term variability appears to be associated with processes within the accretion disc, which appears to be more prominent towards longer optical wavelengths. This variability is likely induced by instabilities within the accretion disc. However, on shorter timescales, particularly on the intra-day timescale, the observed variability may be primarily attributed to jet-related processes. \par
On the other hand, based on the mid-infrared color-magnitude analysis $\emph {(W1-W2)}$ vs. $\emph {W1+W2}$, a large percentage $\sim$ 55\%, $\sim$ 28\%, and $\sim$ 30\%, of gNLS1s, ngNLS1s, and gBLS1s, respectively, exhibit an RWB trend similar to that exhibited in the optical color-magnitude diagram $\emph {(r-i)}$ vs. $\emph {r+i}$ (see Table~\ref{tab:OP_IR_color_variability_stat}). The observed RWB trend in the MIR wavelength in the current sample of Seyfert galaxies can be attributed to the dominance of a variable nuclear component over the constant host galaxy component, where the increased brightness of the AGN leads to a stronger contribution from the AGN dust torus to the observed MIR emission, resulting in a redder $\emph {(W1-W2)}$ color~\citep[e.g., see][]{Stern2012ApJ...753...30S, Yang2018MNRAS.477.5127Y}. Additionally, a similar color trend observed in optical and mid-infrared wavelengths for most of the sources also reconfirms the reprocessing scenario.

\section{Conclusions}
\label{sect_5.0}
We conducted a systematic investigation of flux and color variability in a redshift-matched sample of Seyfert galaxies, comprising 23 gNLS1s, 190 ngNLS1s, and 10 gBLS1s across optical and mid-infrared wavelengths. This analysis utilizes high-cadence photometric observations from the ZTF in the $\emph {g}$, $\emph {r}$, and $\emph {i}$  bands, along with mid-infrared measurements in the $\emph {W1}$ and $\emph {W2}$  bands from the WISE. To ensure the robustness of the spectral analysis, we impose a criterion of at least five quasi-simultaneous observations (within 30 minutes) in the consecutive bands for each source. Most sources have observational baselines that extend up to $\sim$2000 days, allowing a comprehensive study of both long-term flux and color variability in this sample. The primary conclusions drawn from our analysis are as follows:\\
(i) The highest variability is exhibit by gBLS1s in both optical ($\emph {g}$, $\emph {r}$, $\emph {i}$  bands) and mid-infrared ($\emph {W1}$, $\emph {W2}$  bands), followed by gNLS1s, while ngNLS1s show the least variability across both wavelengths. The higher optical and mid-infrared variability observed in gBLS1s compared to gNLS1s suggests a more closely aligned jet toward the observer line of sight in gBLS1s. This could also be due to the comparatively lower accretion rates of gBLS1, which less effectively contaminate the nonthermal Doppler-boosted synchrotron emission of gBLS1s compared to gNLS1s. However, the least variability (DC of $\sim$ 5\%) shown by ngNLS1s among different samples of Seyfert galaxies studied here suggests their long-term variability due to instabilities in the accretion
disc.\par
(ii)  The variability amplitude cut-off suggests that variability in ngNLS1s is primarily driven by accretion disk instabilities, while in gBLS1s and gNLS1s, both accretion disk fluctuations and jet-related processes contribute.\par
(iii)  A similar kind of variability nature resulted from the samples of gNLS1s, ngNLS1s, and gBLS1s in the mid-infrared $\emph {W1}$ and $\emph {W2}$  bands, supporting a reprocessing of optical/UV emission from the central engine of Seyfert galaxies by the dusty torus to the MIR wavelength. The reprocessing scenario was confirmed based on a significant correlation found between optical and mid-infrared variability amplitudes for gNLS1s and gBLS1s in the $\emph {g}$ and $\emph {r}$ bands, while this correlation is found to be generally weak or statistically insignificant in the $\emph {i}$ band and nearly absent in ngNLS1s, except for a notable trend in the $\emph {i}$ band data, which are very scarcely sampled. Therefore, nearly statistically insignificant correlations are found for ngNLS1s, probably due to a combination of a small sample size and significant contamination of host galaxy light in the optical bands, rather than the absence of an intrinsic physical connection between optical and MIR variability. Furthermore, a strong correlation between optical and MIR variability amplitudes across different time scales supports the reprocessing scenario.\par

(iv) Long-term color-magnitude analysis reveals that most gNLS1s, ngNLS1s, and gBLS1s exhibit an RWB trend in both optical (towards longer wavelength side) and MIR wavelengths, strengthening the reprocessing scenario.  Hence, based on flux and color variability analysis, it suggests that the dominant mechanism driving their long-term variability appears to be associated with processes within the accretion disc. This variability is likely induced by instabilities within the accretion disc. However, on shorter timescales, particularly on the intra-day timescale, the observed variability may be primarily attributed to jet-related processes. A similar RWB color trend found for most sources in both optical and mid-infrared wavelengths strongly supports the reprocessing scenario.\par

(v) The present analysis of gNLS1s, ngNLS1s, and gBLS1s highlights the significance of multi-wavelength studies in disentangling the contributions of various emission mechanisms in Seyfert galaxies. The findings underscore the necessity of a more systematic approach involving larger sample sizes and high-cadence, multi-band observations across different timescales to refine our understanding of the physical processes governing AGN variability in future studies.

\section{Acknowledgment}
We acknowledge the support of the National Key R\&D Program of China (2022YFF0503401) and  the National Science Foundation of China (12133001). The work of LCH was supported by the National Science Foundation of China (12233001), the National Key R\&D Program of China (2022YFF0503401) and China Manned Space Program (CMS-CSST-2025-A09). This research utilizes observational data collected with the 48-inch Samuel Oschin Telescope, as part of the Zwicky Transient Facility (ZTF) initiative. The ZTF project is funded by the National Science Foundation through Grant No. AST-2034437 and is a collaborative effort involving institutions such as Caltech, IPAC, the Weizmann Institute of Science, the Oskar Klein Center at Stockholm University, the University of Maryland, DESY and Humboldt University, the TANGO Consortium in Taiwan, the University of Wisconsin–Milwaukee, Trinity College Dublin, Lawrence Livermore, National Laboratory, and IN2P3 in France. The operational management of ZTF is handled by Caltech Optical Observatories (COO), IPAC, and the University of Washington. This study also incorporates data from the Wide Field Infrared Survey Explorer (WISE), a collaborative mission between the University of California, Los Angeles, and NASA’s Jet Propulsion Laboratory at Caltech, supported by the National Aeronautics and Space Administration.

\section*{Data availability}
The optical and mid-infrared data utilized in the current study are publicly available in ZTF DR23 (https://www.ztf.caltech.edu/ztf-public-releases.html) and the final release of the WISE data (https://irsa.ipac.caltech.edu/Missions/wise.html).

\bibliography{main}
\end{document}